\documentclass[journal=jctc,manuscript=article]{achemso}
\usepackage[version=3]{mhchem} 
\usepackage{amsmath,amsfonts,amssymb}
\usepackage{graphicx}
\usepackage{setspace}
\usepackage{tocloft}
\usepackage{orcidlink}
\usepackage{booktabs}
\usepackage{braket}
\usepackage{bm}
\usepackage{titlesec}
\usepackage{fancyhdr}
\usepackage{colortbl}
\usepackage[utf8]{inputenc}
\usepackage{tcolorbox}
\usepackage{xcolor}
\usepackage{booktabs} 
\usepackage{array} 
\usepackage{multirow}
\usepackage{siunitx} 
\usepackage{colortbl} 
\usepackage{xcolor} 
\usepackage{longtable} 
\usepackage{array}
\usepackage{svg}
\usepackage{graphicx}
\usepackage{subcaption}
\usepackage{adjustbox}
\usepackage{booktabs}             
\usepackage{tabularx}              
\usepackage{caption}              

\usepackage{makecell}  
\UseRawInputEncoding

\usepackage{subcaption}

\author{Nikhil Yenugu}
\affiliation{Department of Chemical Sciences, Indian Institute of Science Education and Research (IISER) Kolkata, West Bengal, 741246, India}

\author{Ashwani K Tiwari}
\affiliation{Department of Chemical Sciences, Indian Institute of Science Education and Research (IISER) Kolkata, West Bengal, 741246, India}

\author{Sangita Sen}
\email{sangita.sen@iiserkol.ac.in}
\affiliation{Department of Chemical Sciences and Centre for Advanced Functional Materials, Indian Institute of Science Education and Research (IISER) Kolkata, West Bengal, 741246, India}

\title
  []{A Quantum Mechanical Approach to the Computation of Rovibrational Spectra of Diatomic Molecules in Strong Magnetic Fields}


\begin{document}

\maketitle

\abstract{In the absence of experimental data for molecular spectra in strong magnetic fields, high resolution and reliable computational spectra are required for the interpretation of spectra collected from highly magnetic astrophysical objects. In this paper, we extend the Wilson-Hamiltonian framework, recently implemented and benchmarked by us (\textit{J. Chem. Theory Comput.}, \textbf{21}, 9753 (2025)
), to a general three-dimensional framework suitable for computing the rovibrational spectra of diatomic molecules in strong uniform magnetic fields. The field-dependent electronic and nuclear Hamiltonians capture full non-perturbative coupling between particle motion and the field making the method applicable to all field strengths. The electric quadrupole transition moment integrals for rovibrational transitions in external static magnetic fields are formulated, implemented, and computed to yield spectra which respect the selection rules of the molecule-field system. The spectral changes with increasing field strength such as shifting, splitting, merging, appearance and disappearance of peaks are noted. Contributions from electrons and nuclei are studied individually, as well as in unison to reveal the underlying physics such as stiffening of the bond, emergence of a rotational barrier, field-induced coupling/decoupling of states, and symmetry-breaking in rotational and vibrational states. These results provide the first fully quantum mechanical computational results for the rovibrational signature of $^1\mathrm{H}_{2}$ in extreme magnetic field environments with accuracy suitable for experimental interpretation. The methodology developed herein has direct relevance for high-field spectroscopy and astrochemical modeling, both for providing computational data as well as for understanding the spectral impact of strong magnetic fields on electronic structure and nuclear motion in molecules.}

\section{Introduction}
\label{sec:intro}

Magnetic fields in the universe exhibit an extraordinary range of magnitudes, extending from the exceedingly weak intergalactic fields~\cite{IGFields1} of  $B \approx 10^{-18} - 10^{-15} \; \text{T}$ to the ultra-intense fields~\cite{SMF1,SMF2, SMF5-NS1,SMF6-NS2} exceeding $10^{11} \; \textbf{T}$ that characterize the surfaces of magnetars. In contrast, laboratory-generated magnetic fields typically attain $45 {-} 100$ T in continuous installations and can transiently exceed 1000 T in pulsed or laser-compression experiments~\cite{SMF3-1000T,SMF3b-1200T}. The discovery of mega-gauss (100 T) magnetic fields on white dwarfs~\cite{SMF4-WD1} during the 1970s, followed by the identification of magnetized neutron stars, firmly established that strong magnetic fields are intrinsic to numerous astrophysical objects. Spectroscopic detection of helium~\cite{He1,He2,He3,He4}, molecular hydrogen~\cite{H2-MF}, and heavier atomic species~\cite{He3, HeavierElements2} in such environments has further emphasized the necessity of understanding the structure, stability, and dynamics of atoms and molecules subjected to intense magnetic fields.
In the regime where the magnetic field approaches one atomic unit  $1 \; \text{a.u.} = 2.35 \; \text{x} \; 10^{5} $ T, the magnetic interaction becomes comparable in magnitude to the Coulomb interaction. Under such conditions, novel bonding mechanisms emerge, unconventional molecular ground states appear, and profound field-dependent alterations of spectra are observed, including state reordering, splitting, and the manifestation of previously dark transitions. These effects collectively demonstrate that intense magnetic fields fundamentally reshape the potential energy landscape and chemical bonding paradigms of molecular systems.

The field-induced modifications of the electronic PES lead to dramatic changes in the nuclear motions and consequent modifications of the expected ro-vibrational spectra. Infra-red (IR) ro-vibrational spectroscopy typically plays an important role in the astrochemical detection of molecules, as these transitions are characterized by relatively narrow and well-resolved line structures enabling reliable molecular identification and quantitative abundance analysis. In contrast, molecular electronic spectra in the UV-Vis region are often broadened and congested due to vibronic coupling, predissociation, and overlapping band systems, which can complicate spectral assignment and reduce their diagnostic utility in many stellar environments. Moreover, IR radiation has a greater ability to penetrate intervening material compared with ultraviolet and visible light, making it a key tool for studying the molecular universe. While field-free situations can be modeled with high accuracy using quantum-mechanical rovibrational Hamiltonians, no standard protocols for modeling high magnetic field IR spectra exists. This work aims at developments in this direction by providing a fully quantum protocol to model the ro-vibrational spectra of diatomic molecules in the IR region.

Systematic investigations of molecular systems in strong magnetic fields gained momentum in the 1980s and 1990s, motivated primarily by astrophysical observations of magnetic white dwarfs and neutron stars. Early ab initio studies were largely confined to atomic systems, constrained by limited basis-set flexibility, computational feasibility, and gauge ambiguities in the vector potential corresponding to the field.
Within the Born–-Oppenheimer (BO) approximation, the nuclei are regarded as fixed, while the magnetic field enters the electronic Hamiltonian through both the vector potential and the Zeeman interaction, coupling to the orbital and spin degrees of freedom of the electrons. However, introducing the vector potential into the Hamiltonian through \textit{minimal coupling} presents the challenge of gauge-dependence, which complicates the formulation of non-perturbative methods (a necessity for the strong field regime) for molecular systems in magnetic fields. London atomic orbitals (LAOs) introduced by Fritz London~\cite{London1937} (1937) incorporate an explicit, field-dependent phase factor in the atomic basis associated with the magnetic vector potential, thereby ensuring gauge-origin invariance within a specific gauge function choice (usually the Coulomb gauge) at the orbital level. Subsequent developments by Hameka~\cite{Hameka1} (1958) and Ditchfield~\cite{Ditchfield1, Ditchfield2, Ditchfield3} (1976) extended this concept to gauge-including atomic orbitals (GIAOs), which constitute the formal basis of contemporary electronic-structure methodologies in magnetic fields. 
The introduction of GIAOs~\cite{Helgaker1} in variational, non-perturbative formulations of the electronic structure problem within both Hartree-Fock~\cite{London123-HF3} (HF) and Kohn–-Sham~\cite{CDFT1} (KS) frameworks in the LONDON~\cite{London11, London2,London123-HF3} program enabled the first gauge-invariant and consistent evaluation of atomic and molecular energies and properties in external magnetic fields across the entire regime from weak to ultra-strong. Subsequent methodological developments extended this framework to correlated electronic-structure theories, incorporating electron correlation and magnetically induced bonding phenomena. Implementations have been achieved at all standard levels of theory, including Hartree–Fock~\cite{HF1,HF2}, full configuration interaction~\cite{PB1-FCI,FCI2} (FCI), coupled-cluster~\cite{CC1,CC2,CC3,CC4} (CC), many-body perturbation theory (MBPT), linear-response theory~\cite{LRT1}, the GW approximation~\cite{GWThoery1,GWTheory2}, and density-functional theory~\cite{CDFT1,CDFT2,CDFT3,CDFT4,CDFT5} (DFT). These methodologies have been incorporated in a range of program packages, including LONDON~\cite{London11,London2,London123-HF3}, QUEST~\cite{Quest1}, ChronusQ~\cite{Chronousq1,Chronousq2,Chronousq3}, BAGEL~\cite{Bagel1,Bagel2}, CFOUR~\cite{CFOUR1}, QCUMBRE~\cite{QCUMBRE1,CC2,QCUMBRE3}, TURBOMOLE~\cite{TURBOMOLE1,TURBOMOLE2,TURBOMOLE3} and others~\cite{pyscf1}.

While the electronic structure of atoms and molecules in magnetic fields has been investigated over the last two decades, comparatively little attention has been devoted to molecular dynamics in these environments. The nuclear motion in a magnetic field experiences Lorentz forces that are partially screened by the electronic response leading to a modified set of nuclear equations of motion that couple translational, rotational, and vibrational degrees of freedom. The theoretical foundations of this phenomenon were established more than three decades ago by Schmelcher, Cederbaum, and Meyer~\cite{Schmelcher1,Schmelcher2,Schmelcher3} and Mead and coworkers~\cite{Mead1,Mead2,Mead3}, who formulated the Born - Oppenheimer nuclear dynamics in magnetic fields incorporating Berry-curvature-mediated corrections. However, systematic simulation studies explicitly including these effects have emerged only recently.
The earliest computational approaches to molecular motion in magnetic fields employed a classical perspective. Spreiter and Walter~\cite{CMD1} modified the conventional velocity-Verlet~\cite{VV1} algorithm to include Lorentz forces via a Taylor-series expansion, enabling numerically stable propagation of charged particles in uniform magnetic fields. This formulation was subsequently incorporated into large-scale molecular dynamics packages such as NAMD~\cite{NAMD1} and LAMMPS~\cite{LAMMPS1}, facilitating exploration of magnetically induced molecular motion using empirical force fields. Ceresoli, Marchetti, and Tosatti~\cite{Ceresoli1}, performed the first ab initio molecular dynamics simulations of the $\mathrm{H}_2$ molecule in a perpendicular magnetic field using Hartree-Fock wavefunctions constructed from London orbitals to construct the electronic Berry screening of the Lorentz force.
Building on these insights, Culpitt, Peters, Tellgren, and Helgaker~\cite{AIMD1,AIMD2} developed a rigorous ab initio molecular dynamics formalism that explicitly incorporates the Berry curvature tensor into the Born-Oppenheimer nuclear equations of motion. This framework required the introduction of nonseparable Hamiltonian propagators, most notably the Auxiliary Coordinates and Momenta (ACM) propagator, initially proposed by Tao~\cite{Tao1} and subsequently adapted for Berry-screened Lorentz dynamics. Applications to $\mathrm{H}_2$ and $\mathrm{He}$ revealed pronounced deviations from field-free behavior, including cyclotron-vibrational coupling, hindered rotational motion, and field mediated rovibrational mixing.
More recently, Peters, Tellgren, and Helgaker extended this formalism by introducing the Tajima (TAJ) and Exponential (EXP) propagators~\cite{Propagators1,VDF1}, facilitating longer trajectory simulations and higher-resolution spectral analyses. All these ab initio frameworks thus far, treat electrons quantum mechanically and nuclei classically with the implication that the IR spectra do not reflect the true intensities and selection rules owing to the absence of the knowledge of nuclear wave-functions. The Fourier transform of the velocity autocorrelation function (VACF) has been used in these molecular dynamics simulations to compute the vibrational densities of states and capture anharmonic effects which does not incorporate transition operators such as the dipole moment or quadrupole moment, and therefore fails to enforce spectroscopic selection rules or yield physically meaningful relative intensities. All vibrational modes that are dynamically active will appear, regardless of whether they are IR active or not. 
High-amplitude or low-frequency modes may be artificially dominant with no direct relation to experimental absorption strengths. Furthermore, classical trajectories can easily miss spectral peaks due to insufficient or biased sampling along with spectral changes stemming from quantum tunneling. Extending a fully quantum-mechanical treatment to the nuclear degrees of freedom remains a largely unexplored domain with only a preliminary attempt~\cite{SDNEO1} in the Nuclear-Electronic Orbital (NEO) framework.

Parallel to these developments, efforts have been undertaken to extend beyond the Born-Oppenheimer approximation to incorporate coupled electronic-nuclear dynamics in strong magnetic fields. Although general applications of such non-Born-Oppenheimer approaches remain limited, Adamowicz and co-workers~\cite{Non-BO-1,Non-BO-2,Non-BO-3} have achieved some progress in this direction by studying the HD molecule in a magnetic field using explicitly correlated Gaussian functions beyond the BO approximation. Related progress has also been achieved through the time-dependent nuclear-electronic orbital (TD-NEO) Hartree-Fock framework~\cite{TD-NEO1}, which allows for a coupled quantum-mechanical description of electrons and nuclei. In addition, there have been recent developments of a phase-space electronic Hamiltonian~\cite{PhaseSpace1,PhaseSpace2} that offer us the opportunity to go beyond standard BO theory and perform electronic structure calculations in a moving nuclear frame. Analytical formulations of field-dependent forces and gradients have also been derived, enabling geometry optimization and dynamical studies in magnetic fields~\cite{Opt1, CDFT3, Opt3}. Nevertheless, due to its clarity and computational tractability, the Born-Oppenheimer approximation continues to provide the most robust and widely adopted theoretical framework for the exploration of molecular systems.

In the present study, we remain within the non-relativistic Born-Oppenheimer approximation and employ the Wilson Hamiltonian formalism~\cite{Governale1,jctc1}, originally developed within lattice gauge theory~\cite{WilsonH1,WilsonH2}, to describe the nuclear degrees of freedom quantum mechanically in the presence of magnetic fields. This formalism has the crucial advantage of explicit gauge invariance and is implemented through the Peierls substitution. 
We have recently benchmarked this formalism and demonstrated its application to one electron quantum dots~\cite{jctc1,PCCP2026} and 2D two-body systems~\cite{PCCP2026}.
Here, we extend this approach to a fully three-dimensional grid representation, thereby encompassing all nuclear degrees of freedom and allowing for arbitrary field orientations and potential landscapes. To our knowledge, this constitutes the first fully quantum, gauge-invariant framework for solving the nuclear Schr{\"o}dinger equation of homonuclear diatomic molecules in strong magnetic fields.

The work is organized as follows. Section II presents the theoretical framework, including the Born-Oppenheimer nuclear Hamiltonian and its discretization via the Peierls substitution. Section III describes the computational methodology, encompassing the electronic-structure calculations, interpolation of field-dependent potential energy surfaces, and numerical diagonalization of the Wilson Hamiltonian. Section IV presents and discusses the results, focusing on field-induced energy shifts, rovibrational coupling, and differences between the classical and quantum spectra. Section V concludes the work with perspectives on extending the present framework.

\section{Hamiltonian}
\label{sec:theory}

The non-relativistic Hamiltonian for the hydrogen molecule in an external magnetic field 
$\mathbf{B} = \nabla \times \mathbf{A}$, including spin interactions, using standard symbols is
\begin{equation}
\hat{H} = \hat{T}_e + \hat{T}_N + \hat{V}_{ee} + \hat{V}_{eN} + \hat{V}_{NN} 
+ \hat{H}_{\text{Z},e}^{(\text{spin})} + \hat{H}_{\text{Z},N}^{(\text{spin})},
\label{eq:H_total}
\end{equation} where the components are defined as
\begin{align}
\hat{T}_e &= \sum_{i=1}^{2} 
\frac{1}{2m_e}\left[-i\hbar\nabla_{\mathbf{r}_i} + e\,\mathbf{A}(\mathbf{r}_i)\right]^2,
\label{eq:T_e}\\[4pt]
\hat{T}_N &= \sum_{\alpha=1}^{2} 
\frac{1}{2M_p}\left[-i\hbar\nabla_{\mathbf{R}_\alpha} - q_\alpha\,\mathbf{A}(\mathbf{R}_\alpha)\right]^2,
\label{eq:T_N}\\[4pt]
\hat{V}_{ee} &= \frac{e^2}{|\mathbf{r}_1 - \mathbf{r}_2|},
\label{eq:V_ee}\\[4pt]
\hat{V}_{eN} &= -\sum_{i=1}^{2}\sum_{\alpha=1}^{2}\frac{e^2}{|\mathbf{r}_i - \mathbf{R}_\alpha|},
\label{eq:V_eN}\\[4pt]
\hat{V}_{NN} &= \frac{e^2}{R}, \quad R = |\mathbf{R}_1 - \mathbf{R}_2|,
\label{eq:V_NN}\\[4pt]
\hat{H}_{\text{Z},e}^{(\text{spin})} &= g_e \mu_B \sum_{i=1}^2 \mathbf{S}_i \cdot \mathbf{B},
\label{eq:Zeeman_e}\\[4pt]
\hat{H}_{\text{Z},N}^{(\text{spin})} &= -g_p \mu_N \sum_{\alpha=1}^2 \mathbf{I}_\alpha \cdot \mathbf{B}.
\label{eq:Zeeman_N}
\end{align}
Here $m_e$ and $M_p$ are the electron and proton masses, $\mu_B = e\hbar/(2m_e)$ and 
$\mu_N = e\hbar/(2M_p)$ are the Bohr and nuclear magnetons, respectively.  
For a homogeneous magnetic field $\mathbf{B}$, the vector potentials in the symmetric gauge are given by,
\begin{equation}
\mathbf{A}(\mathbf{r}) = \frac{1}{2}\,\mathbf{B} \times \mathbf{(r-O)} \; \text{and} \; \mathbf{A}(\mathbf{R}) = \frac{1}{2}\,\mathbf{B} \times \mathbf{(R-O)}
\label{eq:symmetric_gauge}
\end{equation}
where O is an arbitary choice of gauge origin.


With nuclear coordinates denoted as $\{\mathbf{R}_\alpha\}_{\alpha=1,2}$, and electronic coordinates denoted as $\{\mathbf{r}_i\}_{i=1,2}$, the total molecular wavefunction for any state under the Born-Oppenheimer approximation can be written as
\begin{equation}
\Psi(\{\mathbf{r}_i\},\{\mathbf{R}_\alpha\}) 
= \psi(\{\mathbf{r}_i\};\{\mathbf{R}_\alpha\})\,\chi(\{\mathbf{R}_\alpha\}),
\label{eq:BO_wavefunction}
\end{equation}
where $\psi(\{\mathbf{r}_i\};\{\mathbf{R}_\alpha\})$ is the electronic wavefunction parameterized by the nuclear positions and $\chi(\{\mathbf{R}_\alpha\})$ is the nuclear wavefunction.

With the electronic Hamiltonian defined as,
\begin{equation}
\hat{H}_e(\{\mathbf{r}_i\};\{\mathbf{R}_\alpha\},\mathbf{B}) 
= \sum_{i=1}^2 \frac{1}{2m_e}\left[-i\hbar\nabla_{\mathbf{r}_i} + e\,\mathbf{A}(\mathbf{r}_i)\right]^2 
+ \hat{V}_{ee} + \hat{V}_{eN}(\{\mathbf{r}_i\};\{\mathbf{R}_\alpha\}) 
+ \frac{e^2}{R} + g_e \mu_B \sum_i \mathbf{S}_i \cdot \mathbf{B},
\label{eq:H_electronic}
\end{equation}
and the nuclear Hamiltonian as,
\begin{equation}
\hat{H}_N = \sum_{\alpha=1}^2 
\frac{1}{2M_\alpha}\left[-i\hbar\nabla_{\mathbf{R}_\alpha} - q_\alpha \,\mathbf{A}(\mathbf{R}_\alpha)\right]^2 
- g_p \mu_N \sum_{\alpha=1}^2 \mathbf{I}_\alpha\cdot\mathbf{B}.
\label{eq:H_nuclear}
\end{equation}
the electronic eigenproblem, under the BO approximation, is given by
\begin{equation}
\hat{H}_e(\{\mathbf{r}_i\};\{\mathbf{R}_\alpha\},\mathbf{B})\,\psi_n(\{\mathbf{r}_i\};\{\mathbf{R}_\alpha\}) 
= \mathcal{E}_n(\{\mathbf{R}_\alpha\})\,\psi_n(\{\mathbf{r}_i\};\{\mathbf{R}_\alpha\}),
\label{eq:electronic_eigen}
\end{equation}
with normalization $\langle\psi_m|\psi_n\rangle=\delta_{mn}$.

Projecting the total Schrodinger equation onto a particular electronic eigenstate, $\psi_k^\ast$, yields the exact nuclear equation with geometric (Berry) coupling:

\begin{equation}
\begin{aligned}
\Bigg[
\sum_{\alpha=1}^2 \frac{1}{2M_\alpha}\Big(-i\hbar\nabla_{\mathbf{R}_\alpha} 
- q_\alpha \,\mathbf{A}(\mathbf{R}_\alpha) 
- \mathbf{A}_{\alpha}^{(k)}(\{\mathbf{R}_\alpha\})\Big)^2 \\
+\, U^{(k)}(\{\mathbf{R}_\alpha\}) 
+ \Phi^{(k)}(\{\mathbf{R}_\alpha\})
\Bigg]\chi_k(\{\mathbf{R}_\alpha\}) = E\,\chi_k(\{\mathbf{R}_\alpha\}),
\end{aligned}
\label{eq:BO_exact_nuclear}
\end{equation}
where
\begin{align}
U^{(k)}(\{\mathbf{R}_\alpha\}) &= \langle \psi_k | \hat{H}_e | \psi_k \rangle_{\mathbf r},
\label{eq:U_BO_explicit}\\[3pt]
\mathbf{A}_{\alpha}^{(k)}(\{\mathbf{R}_\alpha\}) &= -i\hbar\,
\langle \psi_k | \nabla_{\mathbf{R}_\alpha}|\psi_k \rangle_{\mathbf r},
\label{eq:A_geom_alpha}\\[3pt]
\Phi^{(k)}(\{\mathbf{R}_\alpha\}) &= 
\sum_{\alpha=1}^2 \frac{\hbar^2}{2M_\alpha}\sum_{l\neq k}\left|\langle \psi_l | \nabla_{\mathbf{R}_\alpha} \psi_k \rangle_{\mathbf r}\right|^2,
\label{eq:Phi_geom_explicit}\\[3pt]
\mathbf{B}_{\alpha}^{(k)}(\{\mathbf{R}_\alpha\}) &= 
\nabla_{\mathbf{R}_\alpha}\times\mathbf{A}_{\alpha}^{(k)}(\{\mathbf{R}_\alpha\}).
\label{eq:Berry_curvature_alpha}
\end{align}

On neglecting the Berry terms,
\begin{equation}
\Phi^{(k)}(\{\mathbf{R}_\alpha\}) \approx 0, \qquad 
\mathbf{A}_{\alpha}^{(k)}(\{\mathbf{R}_\alpha\}) \approx \mathbf{0}.
\label{eq:BO_approx_alpha}
\end{equation}
we arrive at the nuclear Hamiltonian on the $k$-th Born-Oppenheimer potential energy surface $U^{(k)}$ used by us,
\begin{equation}
\hat{H}_N^{(k)} \;=\; 
\sum_{\alpha=1}^2 \frac{1}{2M_\alpha}\left[-i\hbar\nabla_{\mathbf{R}_\alpha} - q_\alpha \,\mathbf{A}(\mathbf{R}_\alpha)\right]^2 
+ U^{(k)}(\mathbf{R}) \;-\; g_p \mu_N \sum_{\alpha=1}^2 \mathbf{I}_\alpha \cdot \mathbf{B},
\label{eq:H_N_final_alpha}
\end{equation}
where $U^{(k)}(\mathbf{R})$ depends implicitly on the internuclear distance $R = |\mathbf{R}_1 - \mathbf{R}_2|$.
We neglect the nuclear spin term in this paper but it is straightforward to include as and when required. The superscript `k' is also dropped as we restrict our discussions to the lowest singlet electronic state of H$_2$ at all fields corresponding to the zero-field molecular state, $^1\Sigma_{g}^+$.

\subsection{Separation of Centre of Mass Motion} \label{sec:COM}

The nuclear Hamiltonian of Eq.~\ref{eq:H_N_final_alpha} for the two-body $H_2$ molecule involves the masses $ M_1, M_2 $ and charges $q_1, q_2$, of the two nuclei, their positions $ \mathbf{R}_{\textnormal{1}} $ and $ \mathbf{R}_{\textnormal{2}} $, and their momenta $ \mathbf{P}_{\textnormal{1}} $ and $ \mathbf{P}_{\textnormal{2}} $. 
In an external magnetic field, $\mathbf{\hat{B}}=\mathbf{\hat{\nabla}}\times\mathbf{\hat{A}}(\mathbf{R}_\alpha) $, the particles experience vector potentials, $\mathbf{\hat{A}}_\mathrm{1} = \mathbf{\hat{A}}(\mathbf{R}_\mathrm{1})$ and $\mathbf{\hat{A}}_\mathrm{2} = \mathbf{\hat{A}}(\mathbf{R}_\mathrm{2})$ respectively, leading to a simplified notation for the Hamiltonian of Eq.~\ref{eq:H_N_final_alpha} as,
\begin{equation}
    \hat{\mathrm{H}}_{\mathrm{N}} = \frac{{(\mathbf{\hat{P}}_1-q_1\mathbf{\hat{A}}_1)^2}} {{2M_1}} + \frac{{(\mathbf{\hat{P}}_2- q_2\mathbf{\hat{A}}_2)^2}} {{2M_2}} + \hat{V}(|\mathbf{R}_1-\mathbf{R}_2|) \label{hamOwithB}
\end{equation}

In our earlier publication~\cite{jctc1,PCCP2026} we have reduced this to an effective one-body problem in the COM frame through the introduction of the reduced-mass $ \mu_M = \frac{{M_1 M_2}}{{M_1 + M_2}} $, and a set of reduced mass-weighted charges,
\begin{equation}
    q_{R_{CM}} = q_1 + q_2
\end{equation}

\begin{equation}
    q_R = \frac{M_2^2q_1 + M_1^2q_2}{M^2}
    \label{q_r}
\end{equation}

\begin{equation}
    q_{RR_{CM}} = \frac{M_2q_1 - M_1q_2}{M}
\end{equation}
and denoting the vector potentials in the COM and relative coordinates as $\mathbf{\hat{A}_R}$ and $\mathbf{\hat{A}_r}$ respectively.
The Hamiltonian, in atomic units, represented in the center-of-mass (COM) coordinates $\mathbf{R_{CM}} = \frac{{M_1 \mathbf{R}_1 + M_2 \mathbf{R}_2}}{{M_1 + M_2}} $ and the relative coordinate $\mathbf{R}=\mathbf{R}_{\textnormal{1}} - \mathbf{R}_{\textnormal{2}}$ then assumes the form,

\begin{equation}
\hat{\mathrm{H}}_{N}=\frac{\left(\hat{\mathbf{P}}_\mathrm{R}-q_{\mathbf{R}} \hat{\mathbf{A}}_{\mathbf{R}}-q_{RR_{CM}} \hat{\mathbf{A}}_{\mathbf{R_{CM}}}\right)^2}{2 \mu_M}+\frac{\left(\hat{\mathbf{P}}_\mathrm{R_{CM}}-q_{R_{CM}} \hat{\mathbf{A}}_{\mathbf{R_{CM}}}-q_{RR_{CM}} \hat{\mathbf{A}}_{\mathbf{R}}\right)^2}{2 M}+V(\mathbf{R})
\label{qw}
\end{equation}

This Hamiltonian allows us to partition the COM motion and relative motion and identify the coupling terms as follows:
\begin{equation}
\hat{\mathrm{H}}_R = \frac{\left(\hat{\mathbf{P}}_\mathrm{R} - q_R \hat{\mathbf{A}}_{\mathbf{R}}\right)^2}{2 \mu_M} + V(\mathbf{R}), 
\label{eq:Hr}
\end{equation}
\begin{equation}
\hat{\mathrm{H}}_{R_{CM}} = \frac{\left(\hat{\mathbf{P}}_\mathrm{R_{CM}} - {q_{R_{CM}}} \hat{\mathbf{A}}_{\mathbf{R}}\right)^2}{2 M}, 
\label{eq:HR}
\end{equation}
\begin{equation}
\begin{aligned}
\hat{\mathrm{H}}_{R R_{CM}} &= -q_{R R_{CM}} \left( \frac{1}{\mu_M} \hat{\mathbf{A}}_{\mathbf{R_{CM}}} \hat{\mathbf{P}}_\mathrm{R_{CM}} 
+ \frac{1}{M} \hat{\mathbf{A}}_{\mathbf{R}} \hat{\mathbf{P}}_\mathrm{R_{CM}} \right) \\
&\quad + {q_{R R_{CM}}^2} \left[ \left(\frac{1}{M} + \frac{1}{\mu_M}\right) \hat{\mathbf{A}}_{\mathbf{R}_{CM}} \hat{\mathbf{A}}_{\mathbf{R}} 
+ \frac{1}{2 \mu_M} \hat{\mathbf{A}}_{\mathbf{R_{CM}}}^2 + \frac{1}{2 M} \hat{\mathbf{A}}_{\mathbf{R}}^2 \right].
\end{aligned}
\label{eq:HrR}
\end{equation}
with
\begin{equation}
    \mathrm{\hat{H}}_{N} = \mathrm{\hat{H}}_{R_{CM}} + \mathrm{\hat{H}}_R + \mathrm{\hat{H}}_{RR_{CM}}
\end{equation}
Since $q_{RR_{CM}}$ is multiplied to all the terms in $H_{RR_{CM}}$ and $q_{RR_{CM}}=0$ for H$_2$ where both nuclei have the same charge-mass ratio, ie. $\frac{q_1}{M_1} = \frac{q_2}{M_2}$, we can shift to the effective one-body reduced-mass Hamiltonian in the COM frame without any approximation, leading to,
\begin{equation} \label{Hone}
    \mathrm{\hat{H}}_{N} = \mathrm{\hat{H}}_R 
\end{equation}.

\subsection{Discretization of the Nuclear Hamiltonian}
\label{sec:Ham}
The effective one-particle nuclear Hamiltonian in COM frame $\hat{H}_R$ of Eq.~\ref{Hone} can be discretized on a grid and solved in the gauge-invariant Wilson Hamiltonian framework recently implemented by us~\cite{jctc1,PCCP2026}. The formalism and implementation have been benchmarked therein against the Fock-Darwin eigenstates and eigenfunctions for the 2D harmonic oscillator (HO) model in a perpendicular magnetic field. This exercise also helped us establish the protocol for deciding grid parameters and obtaining the optimal grid parameters for the $H_2$ molecule.
For treating the complete rovibrational problem, we now extend the formalism and implementation to the 3D situation where all orientations of the molecule relative to the magnetic field can be considered. The electronic PES, $V(\textbf{R}) = V(R,\theta)$ is a function of the internuclear distance, $R$, and the angle between the bond and the magnetic field (considered to be along the z-axis), $\theta$. The potential $V(R,\theta)$ can be rotated about the z-axis to generate the 3D PES $V(R,\theta,\phi)$ where $\phi$ is the angle between the projection of $\textbf{R}$ on the xy-plane and the y-axis.

On the uniform 3D grid any generic point is represented as $\mathbf{R}=(c_x\Delta \text{x}, c_y\Delta \text{y}, c_z\Delta \text{z})$,
$c_x$, $c_y$, and $c_z$ being integers and $\Delta$x, $\Delta$y and $\Delta$z being the grid-spacings along the x, y and z axes respectively. Any generic direction in the discretized space is denoted as $\eta$ and $a_\eta$ indicates the grid-spacing along that direction. A lattice operator $U_\eta (\mathbf{R})$ can be defined as:
\begin{equation} \label{eq:unitary_operator}
    U_\eta(\mathbf{R}) = \text{e}^{iga_\eta A_\eta(\mathbf{R})} \; \; \; ; \; \eta = \mathrm{x},\mathrm{y}, \mathrm{z} 
\end{equation}
The discretized Wilson Hamiltonian on a 3D grid for any two grid points $(\mathrm{x}_1,\mathrm{y}_1, \mathrm{z}_1) \; \text{and} \; (\mathrm{x}_2,\mathrm{y}_2, \mathrm{z}_2)$ of Eq.~\ref{eq:unitary_operator} can then be written as,
\begin{equation}\label{Discrete_H}
\begin{gathered}
\hat{H}_{\mathrm{WH}}(x_{1}, y_{1}, z_{1}, x_{2}, y_{2}, z_{2}) = 
\sum_{\eta} \frac{\hbar^2}{2 M_{\eta}} 
\Bigg[
\frac{\delta_{y_{1}, y_{2}} \, \delta_{z_{1}, z_{2}}}{(\Delta x)^2}
\Big( 2 \delta_{x_{1}, x_{2}} 
- U_{\eta}(x_{1}) \, \delta_{x_{1}-1, x_{2}} 
- U_{\eta}^{\dagger}(x_{2}) \, \delta_{x_{1}+1, x_{2}} \Big) \\[6pt]
+ \frac{\delta_{x_{1}, x_{2}} \, \delta_{z_{1}, z_{2}}}{(\Delta y)^2} 
\Big( 2 \delta_{y_{1}, y_{2}} 
- U_{\eta}(y_{1}) \, \delta_{y_{1}-1, y_{2}} 
- U_{\eta}^{\dagger}(y_{2}) \, \delta_{y_{1}+1, y_{2}} \Big) \\[6pt]
+ \frac{\delta_{x_{1}, x_{2}} \, \delta_{y_{1}, y_{2}}}{(\Delta z)^2} 
\Big( 2 \delta_{z_{1}, z_{2}} 
- U_{\eta}(z_{1}) \, \delta_{z_{1}-1, z_{2}} 
- U_{\eta}^{\dagger}(z_{2}) \, \delta_{z_{1}+1, z_{2}} \Big)
\Bigg] \\[6pt]
+ \hat{V}(x_{1}, y_{1}, z_{1}) \,
\delta_{x_{1}, x_{2}} \,
\delta_{y_{1}, y_{2}} \,
\delta_{z_{1}, z_{2}}.
\end{gathered}
\end{equation}
$U_\eta(\mathbf{R})$ is commonly called the Peierls phase\cite{Peierls1933} and the procedure described above is called the Peierls substitution. Here, we are using a 2nd-order (central) finite difference scheme. The Hamiltonian, $\mathrm{\hat{H}_{WH}}$, is referred to as the Wilson Hamiltonian in the rest of this paper. The gauge invariance of the time independent and time dependent Schrodinger equation has been proved and numerically demonstrated in earlier publications~\cite{Governale1,jctc1,PCCP2026}.

Setting the z-coordinate to zero, the corresponding 2D Wilson Hamiltonian can be written as,
\begin{equation}\label{Discrete_H2D}
\begin{gathered}
\hat{H}_{\mathrm{WH}}(x_{1}, y_{1}, x_{2}, y_{2}) = 
\sum_{\eta} \frac{\hbar^2}{2 M_{\eta}} 
\Bigg[
\frac{\delta_{y_{1}, y_{2}}}{(\Delta x)^2}
\Big( 2 \delta_{x_{1}, x_{2}} 
- U_{\eta}(x_{1}) \, \delta_{x_{1}-1, x_{2}} 
- U_{\eta}^{\dagger}(x_{2}) \, \delta_{x_{1}+1, x_{2}} \Big) \\[6pt]
+ \frac{\delta_{x_{1}, x_{2}}}{(\Delta y)^2} 
\Big( 2 \delta_{y_{1}, y_{2}} 
- U_{\eta}(y_{1}) \, \delta_{y_{1}-1, y_{2}} 
- U_{\eta}^{\dagger}(y_{2}) \, \delta_{y_{1}+1, y_{2}} \Big) \Bigg] \\[6pt]
+ \hat{V}(x_{1}, y_{1}) \,
\delta_{x_{1}, x_{2}} \,
\delta_{y_{1}, y_{2}}
\end{gathered}
\end{equation}
where the wavefunction is confined to the xy-plane and the magnetic field is along the z-axis. The electronic PES $V(x,y)$ is then equivalent to $V(R,\theta=90^\circ,\phi)$. For all numerical calculations, the gauge origin is chosen by default at $(0,0,0)$, unless stated otherwise. This choice does not affect the computed results, as the transition moments and oscillator strengths are invariant with respect to the selection of the gauge origin and the gauge function.

\section{Computation of Rovibrational Spectra}
\label{sec:CD}
The electronic Hamiltonian was solved using the \textsc{London} quantum chemistry program to obtain the electronic potential energy surface of the hydrogen molecule in the presence of an external unifrom magnetic field. Calculations were performed for internuclear separations $R$ ranging from $0.1$ to $10~\mathrm{a.u.}$, and for orientations where the angle between the internuclear axis and the magnetic field direction varied from $\theta = 0^{\circ}$ to $90^{\circ}$ in increments of $5^{\circ}$. For each $(R, \theta)$ configuration, single-point electronic energies were computed at the CCSD level of theory using the Lu-aug-cc-pVTZ basis set where L stands for London Atomic Orbitals (also called Gauge Including Atomic Orbitals, GIAO) which ensure gauge origin invariance of the electronic energy and u stands for `uncontracted'. The resulting electronic energies were subsequently interpolated using a bivariate spline procedure to generate a smooth, continuous representation of the potential energy surface. The fitted surface is supplied in the SI (as a Python code). This interpolated surface was then mapped onto a three-dimensional uniform Cartesian grid, providing the electronic potential for the nuclear Hamiltonian.

The nuclear Hamiltonian was then constructed according to Eq.~\ref{Discrete_H} and the Hamiltonian matrix was diagonalzied using sparse matrix diagonalization (Sparse eigenvalue problems were solved using the implicitly restarted Lanczos method as implemented in ARPACK via SciPy's \textit{eigsh} routine.)
\ in an in-house code. Grid spacings and box-sizes were explored to ensure energy convergence less than ~$10 \;\mathrm{cm}^{-1}$.

To systematically analyze magnetic field effects, three distinct scenarios were considered: (i) the magnetic field acts only on the electrons (i.e., affecting the electronic potential while neglecting nuclear coupling); (ii) the magnetic field acts only on the nuclei (modifying the kinetic operator via the Peierls substitution while keeping the electronic potential field-free); and (iii) the magnetic field acts simultaneously on both electrons and nuclei. This separation allows for direct assessment of the individual and combined contributions of the electronic and nuclear magnetic couplings.

Furthermore, to investigate the role of anharmonicity in the rovibrational spectra, the same three scenarios were repeated using a harmonic potential in place of the interpolated electronic potential. The rovibrational spectra were then computed from the eigenvalues and eigenvectors obtained after solving the nuclear Hamiltonian in all cases, and the resulting comparisons are presented and discussed in Sec.~\ref{sec:results}. 

As the $H_2$ molecule has no permanent electronic dipole moment it shows no dipole-allowed rovibrational transitions at $\textbf{B}=0$. Weak electric quadrupole transitions can be observed. The point group of the molecule+field system is modified in the presence of magnetic fields in various orientations~\cite{Pausch2021MolecularFields}. $H_2$ at $\textbf{B}=0$ belongs to $D_{\infty h}$, with $\textbf{B} \parallel bond$ to $C_{\infty h}$, $\textbf{B} \perp bond$ to $C_{2h}$ and with arbitrary $\textbf{B}$ to $C_i$. However, in all cases the centre of inversion, i, is present and no permanent electronic dipole moment is generated. Thus, the primary contribution to the the transition moment integral (TMI) comes from the electric quadrupole moment.

\subsection{Electric Quadrupole Transition Moment Integral}
\label{sec:QDM}

The electric quadrupole moment represents the second-order contribution in the multipole expansion of the electrostatic potential generated by a charge distribution.\cite{Stone2013,Jackson1999,Buckingham1967} 
In Cartesian coordinates the quadrupole tensor for a system of charges can be written as

\begin{equation}
Q_{pq} = \sum_a q_a \left( 3 r_{a,p} r_{a,q} - r_a^2 \delta_{pq} \right),
\label{eq:quadrupole_general}
\end{equation}

where $q_a$ denotes the charge of particle $a$, $r_{a,i}$ is the $i$-th Cartesian coordinate of that particle, and $\delta_{ij}$ is the Kronecker delta. 
For molecular systems under the BO approximation, the quadrupole tensor can be separated into electronic and nuclear contributions,

\begin{equation}
\hat{Q}_{pq} = \hat{Q}_{pq}^{(e)}(\textbf{r};\textbf{R}) + \hat{Q}_{pq}^{(n)}(\textbf{R}),
\label{eq:quadrupole_decomposition}
\end{equation}
where $\textbf{r}$ and $\textbf{R}$ denote the electronic coordinates and nuclear coordinates, respectively.

In the present work the electronic and nuclear contributions are evaluated using complementary computational approaches. The electronic quadrupole tensor is obtained from basis-set electronic structure calculations, while the nuclear contribution is evaluated using grid-based nuclear wavefunctions. The electronic quadrupole tensor components are obtained using the \textsc{London} electronic structure program, which employs a basis-set expansion incorporating London orbitals appropriate for molecular systems in magnetic fields.\cite{London1937}
For a given internuclear separation $\textbf{R}$, the electronic quadrupole tensor is obtained as the expectation value

\begin{equation}
Q_{pq}^{(e)}(\textbf{R}) =
- e \int \rho(\mathbf{r};\textbf{R})
\left( 3 r_p r_q - r^2 \delta_{pq} \right) d^3 r ,
\label{eq:electronic_quadrupole}
\end{equation}

where $\rho(\mathbf{r};\textbf{R})$ is the electronic charge density.

Because these quantities are computed only for discrete nuclear geometries, the resulting quadrupole tensor components are interpolated using cubic spline interpolation,

\begin{equation}
Q_{pq}^{(e)}(\textbf{R}) \approx S_{pq}(\textbf{R}),
\label{eq:spline}
\end{equation}

where $S_{pq}(\textbf{R})$ denotes the spline representation constructed from the electronic structure data. This procedure provides a smooth and numerically stable representation of the electronic quadrupole tensor across the range of nuclear configurations explored in the nuclear dynamics calculations.

The nuclear Schr\"odinger equation is solved on a uniform Cartesian grid. The nuclear wavefunction is represented as

\begin{equation}
\Psi(x_i,y_j,z_k),
\label{eq:wavefunction_grid}
\end{equation}

where $(x_i,y_j,z_k)$ denote discrete grid coordinates. 
Within this grid representation, the quadrupole operator is evaluated directly from the Cartesian coordinates of the grid points. 
For a point charge ($q_R$ in our case) located at position $\mathbf{R} = (x,y,z)$ the quadrupole operator is
\begin{equation}
Q_{pq}^{(n)}(\mathbf{R}) =
q_R \left( 3 R_p R_q - R^2 \delta_{pq} \right),
\label{eq:nuclear_quadrupole}
\end{equation}
with
\begin{equation}
R^2 = x^2 + y^2 + z^2.
\label{eq:r2}
\end{equation}

The Cartesian tensor components are therefore
\begin{align}
Q_{xx} &= 3x^2 - R^2, \\
Q_{yy} &= 3y^2 - R^2, \\
Q_{zz} &= 3z^2 - R^2, \\
Q_{xy} &= 3xy, \\
Q_{xz} &= 3xz, \\
Q_{yz} &= 3yz .
\label{eq:quadrupole_components}
\end{align}
These expressions are evaluated at each grid point to obtain the nuclear contribution to the quadrupole tensor.

For the reduced-dimensionality (2D) calculations where the nuclear motion is restricted to the $xy$ plane, the quadrupole tensor reduces to
\begin{equation}
Q =
\begin{pmatrix}
Q_{xx} & Q_{xy} \\
Q_{xy} & Q_{yy}
\end{pmatrix},
\label{eq:quadrupole_2d}
\end{equation}
with
\begin{equation}
R^2 = x^2 + y^2 .
\end{equation}
For the full three-dimensional calculations, the quadrupole tensor retains its complete form,
\begin{equation}
Q =
\begin{pmatrix}
Q_{xx} & Q_{xy} & Q_{xz} \\
Q_{xy} & Q_{yy} & Q_{yz} \\
Q_{xz} & Q_{yz} & Q_{zz}
\end{pmatrix}.
\label{eq:quadrupole_3d}
\end{equation}

The expectation value of the quadrupole tensor for a nuclear eigenstate $\Psi_n(\textbf{R})$ is given by
\begin{equation}
\langle Q_{pq} \rangle =
\int
\Psi_n^*(\mathbf{R})
Q_{pq}(\mathbf{R})
\Psi_n(\mathbf{R}) d\mathbf{R}.
\label{eq:quadrupole_expectation}
\end{equation}
where $Q_{pq}(\mathbf{R})=Q^{(e)}_{pq}(\mathbf{R})+Q^{(n)}_{pq}(\mathbf{R})$.
Within the grid representation, the integral is approximated numerically as
\begin{equation}
\langle Q_{pq} \rangle
\approx
\sum_{i,j,k}
\Psi^*(x_i,y_j,z_k)
Q_{pq}(x_i,y_j,z_k)
\Psi(x_i,y_j,z_k)
\Delta V,
\label{eq:grid_integral}
\end{equation}
where
\begin{equation}
\Delta V = \Delta x \Delta y \Delta z
\end{equation}
is the volume element of the grid. 

The quadrupole transition moment integrals between the nuclear eigenstates $m$ and $n$ are calculated as
\begin{equation}
Q_{pq}^{mn} =
\int
\Psi_m^*(\mathbf{R})
Q_{pq}(\mathbf{R})
\Psi_n(\mathbf{R})
d\mathbf{R}.
\label{eq:quadrupole_transition}
\end{equation}

A scalar quadrupole strength is defined as
\begin{equation}
Q^{mn} =
\sqrt{
\frac12 \Bigg[(Q_{xx}^{mn})^2 + (Q_{yy}^{mn})^2 + (Q_{zz}^{mn})^2\Bigg]
+ (Q_{xy}^{mn})^2 + (Q_{xz}^{mn})^2 + (Q_{yz}^{mn})^2 }
\label{eq:quadrupole_scalar}
\end{equation}
which reduces to 
\begin{equation}
Q^{mn} =
\sqrt{\tfrac12 \Bigg[(Q_{xx}^{mn})^2 + (Q_{yy}^{mn})^2\Bigg] + (Q_{xy}^{mn})^2 }.
\label{eq:quadrupole_scalar_2d}
\end{equation}
for the two-dimensional calculations. Giving the electric quadrupole oscillator strength, 
\begin{equation}
f_Q =
\frac{2}{15}
\frac{\omega^5}{c^5}
\left|
Q^{mn}
\right|^2 
\label{eq:quadrupole_osc}
\end{equation}

where $\omega$ or $\hbar \omega \;(\hbar=1)$  is the transition energy and $c$ is the speed of light in atomic units. 

\section{Results and Discussion} \label{sec:results}

The rovibrational spectroscopy of the hydrogen molecule has long served as a fundamental benchmark for molecular quantum theory. In the absence of external fields, the singlet ground state of H$_2$ is accurately described by a separable rigid-rotor and harmonic-oscillator model, with nearly free rotational motion and a dominant vibrational stretching frequency in the range of 4400-4600~cm$^{-1}$. Strong magnetic fields qualitatively and quantitatively reshape this familiar rovibrational spectrum of singlet H$_2$, through gauge-field-induced couplings of electron-electron, electron-nucleus and nucleus-nucleus relative motions~\cite{AIMD2,VDF1,MDLM-1}. A central and unifying result is that the dominant magnetic-field effect on rovibrational spectra originates from the modifications of the Born-Oppenheimer potential energy surface (PES) while the direct Lorentz forces acting on the nuclei induce peak splittings. A few results of only vibrational energy levels in a parallel magnetic field have also been computed through numerical solutions of the radial Schrodinger equation~\cite{pyscf1} where the field has only been considered to act on the electrons as it does not couple to the nuclear vibration.

In this work, a fully quantum treatment of the nuclear motion in the COM frame has been carried out under the BO approximation as discussed in Sec.\ref{sec:theory}. The magnetic vector potential is included non-perturbatively ensuring complete gauge invariance. This formalism is expected to avoid missing peaks, properly account for intensities based on wavefunction overlaps and orthogonality and other exclusively quantum couplings (if any) as opposed to earlier classical MD simulations. This Hamiltonian - based method also avoids bias from initial conditions which is a common shortcoming of trajectory-based methods. The electric quadrupolar rovibrational spectra are then computed for various situations as discussed in Sec.~\ref{sec:CD}.

A quantitative description of these effects necessitates the construction of magnetic-field-dependent PESs. Ref.~\citenum{AIMD2} used PESs constructed at the Hatree-Fock (HF)/Lu-cc-pVDZ level, Ref.~\citenum{MDLM-1} used PESs at the MP2/def2-TZVP level and Ref.~\citenum{pyscf1} at the CCSD/cc-pVTZ level. We have constructed CCSD/Lu-aug-cc-pVTZ PESs which have essentially full CI (FCI) level accuracy for two electronic systems like $^1\mathrm{H}_2$.

\subsection{Potential Energy Surface of $\mathrm{H}_2$ molecule}
\label{subsection:PES}

\begin{figure}[htbp]
    \centering
    \begin{minipage}{0.5\linewidth}
        \centering
        \includegraphics[width=\linewidth]{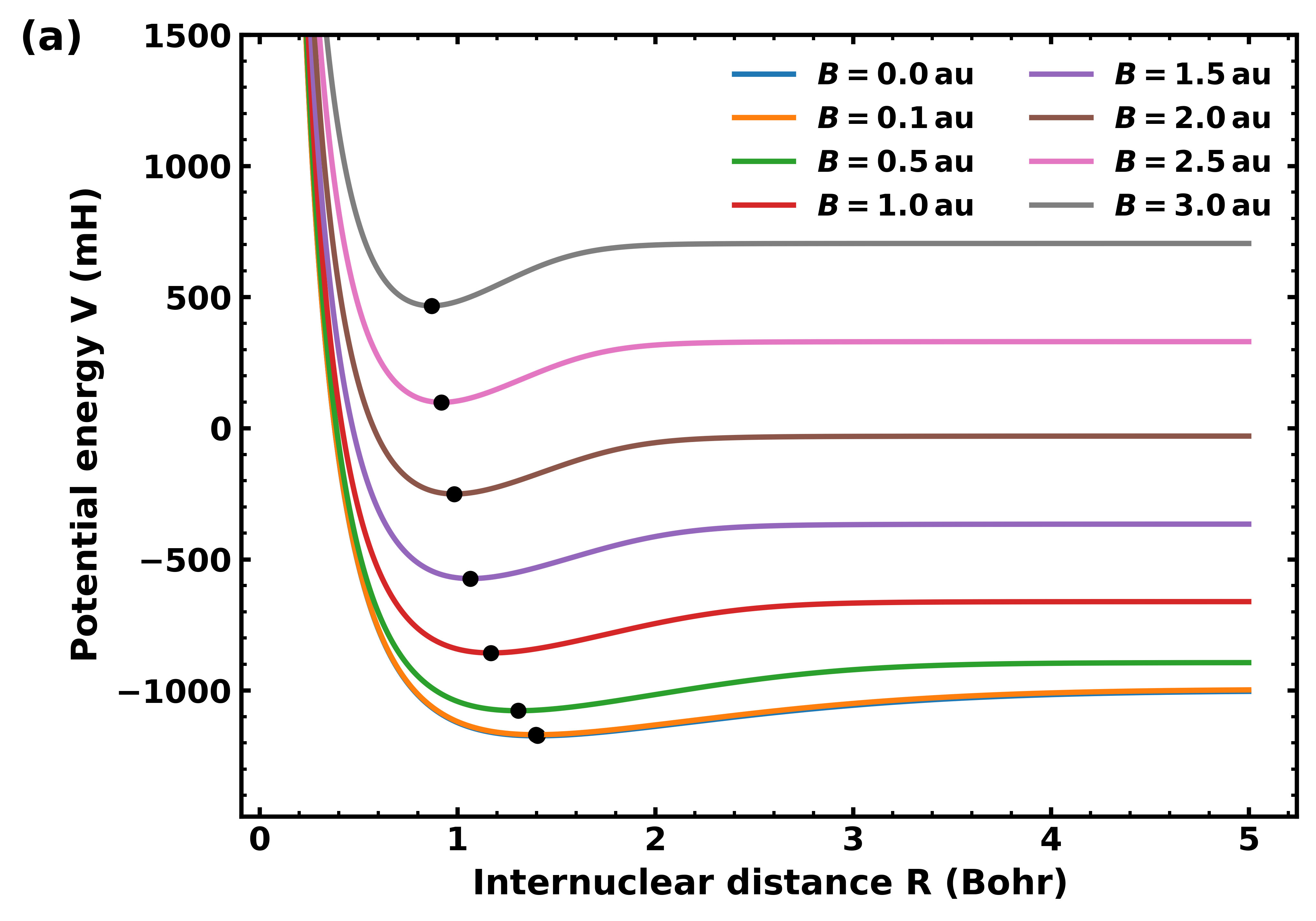}
    \end{minipage}
    \hfill
    \begin{minipage}{0.5\linewidth}
        \centering
        \includegraphics[width=\linewidth]{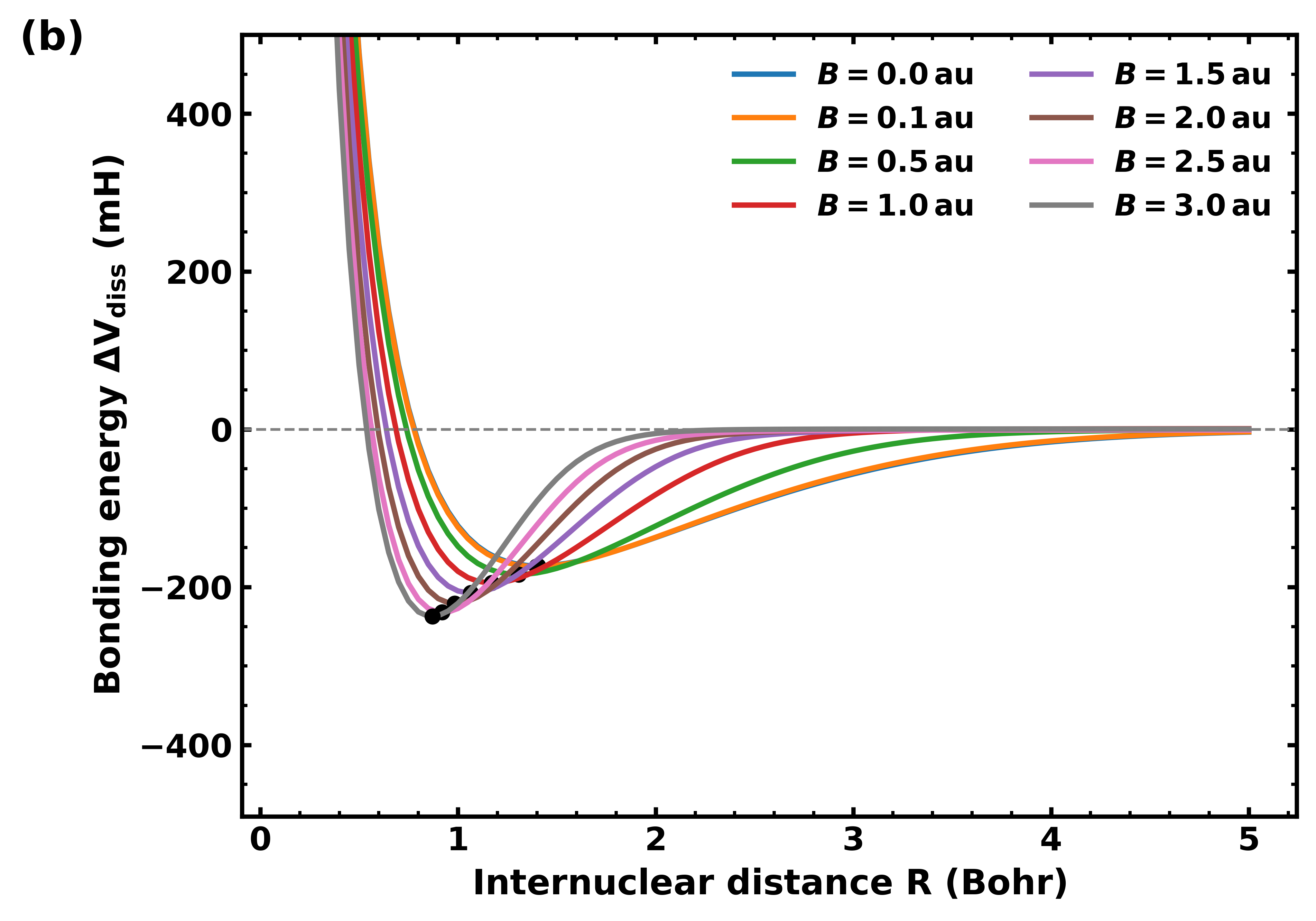}
    \end{minipage}
    \vspace{0.5cm}
    \begin{minipage}{0.5\linewidth}
        \centering
        \includegraphics[width=\linewidth]{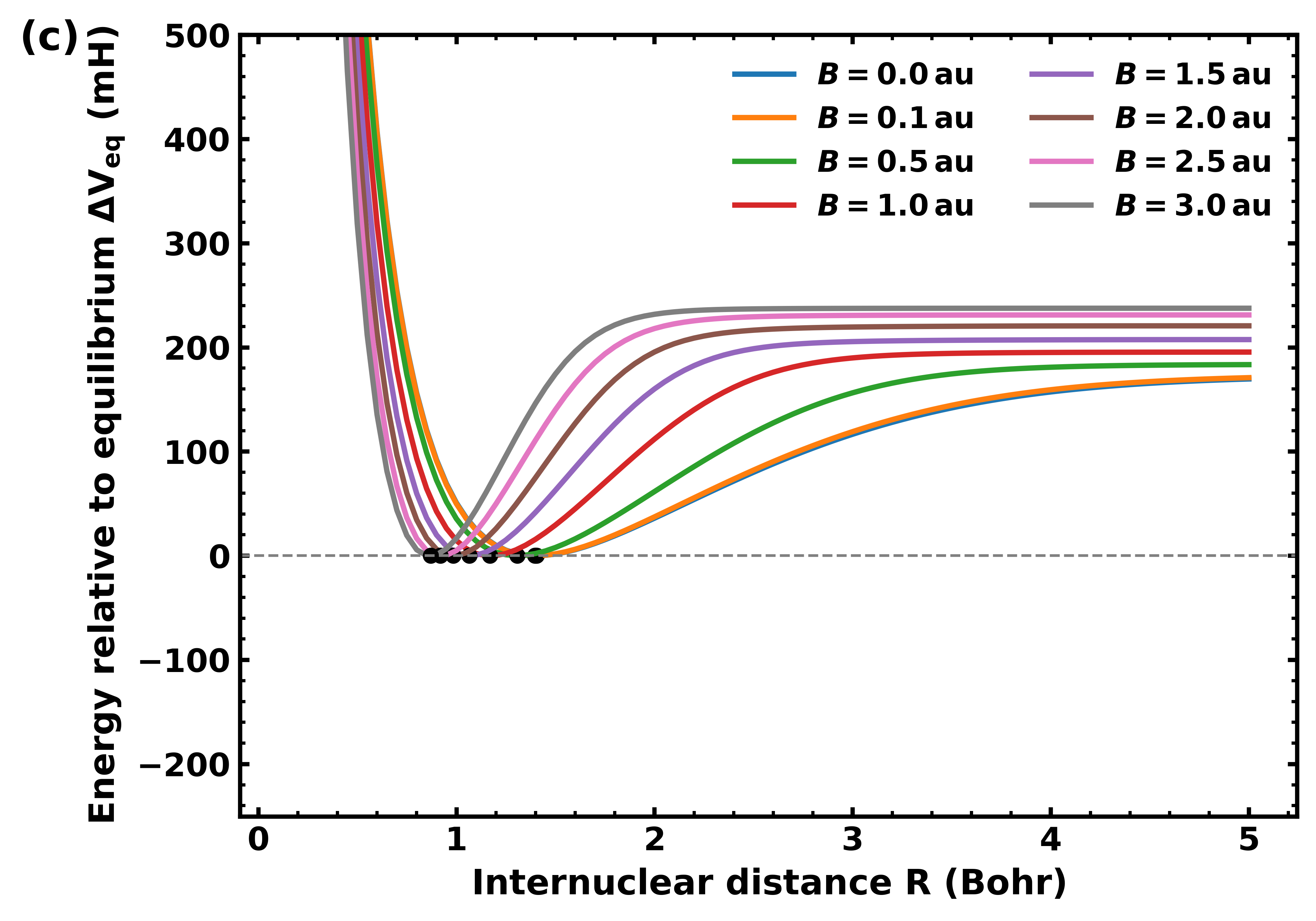}
    \end{minipage}
\caption{
(a) Potential energy curves (PECs) of $^1$H$_2$ for magnetic field strengths $B=0-3$ au, computed at the CCSD/Lu-aug-cc-pVTZ level with the molecular axis perpendicular to the field ($\theta=90^\circ$).
(b) Bonding energy relative to dissociation, $V - V_{\mathrm{diss}}$, for the same field strengths.
(c) Energy relative to the equilibrium minimum, $V - V_{\mathrm{eq}}$, highlighting the field-dependent modification of the molecular bond.
}
    \label{fig:2D-PES1}
\end{figure}

For analyzing the PES, as also described in Sec.~\ref{sec:theory}, we have divided it into two subparts: i) A 2D PES $\mathrm{V}(R,\theta=90^{o},\phi)$ (where the molecular bond is perpendicular to the magnetic field axis and the bond-length $R$ is varied and ii) a 3D PES $\mathrm{V}(R,\theta,\phi)$, where the molecular bond orientation with respect to the magnetic field axis ($\theta$) and the bond length ($R$) are both varied.

\textbf{2D PES:} A 2D PES is generated by rotating the $E_{el} \; \mathrm{vs} \; R$ (say along $\mathrm{x}$-axis) PEC about the $\mathrm{y}$-axis. The molecule is confined to the $\mathrm{xy}$-plane and the magnetic-field is along $\mathrm{z}$-axis. In Fig.~\ref{fig:2D-PES1}a PECs of $H_2$ at CCSD/L-aug-cc-pVTZ are presented PECs with field more strengths are presented in the supplementary information (SI) Fig. S1. It can be noticed that as the magnetic field is increased from $0 \;\mathrm{au} \;\text{to} \;3\; \mathrm{au}$ the potential energy well is shifted up by 1.641 Hartree (H) (44.65 eV), and the equilibrium bond distance is decreased by $\approx 38.03\%$ (See SI Table S1). It is noticeable from Fig.~\ref{fig:2D-PES1}b and Fig.~\ref{fig:2D-PES1}c that the bonding energy is increased by $68\;\mathrm{mH} \; (1.85\; \mathrm{eV})$and the curvature is increased (force constant up by $\approx 84\%$) as the field is increased.

The influence of electron correlation on the magnetic-field-modified $^1\mathrm{H}_2$ potential energy curve is evident through systematic changes in structural, energetic, and spectroscopic parameters from HF to MP2 and CCSD (see SI Table S4), with correlated methods consistently predicting weaker confinement, softer curvature, and enhanced anharmonicity across all fields. At zero field, the HF equilibrium bond length (1.387 au) is shorter than MP2 (1.393 au) and CCSD (1.404 au) by about 0.4\% and 1.2\%, a hierarchy that persists at strong fields - for example, at $B=3.0$~a.u. where HF (0.857 au) remains approximately 1.0\% shorter than MP2 (0.861 au) and nearly 1.1\% shorter than CCSD (0.866 au), indicating that correlation counteracts magnetic compression by stabilizing a more delocalized electronic distribution. Energetic effects are stronger: at zero field, the dissociation energy $D_e$ drops from 0.274~H (HF) to 0.229~H (MP2) and 0.170~H (CCSD), showing that HF overbinds by roughly 20-40\%, while at $B=2.0$~a.u. HF predicts $D_e \approx 0.530$~H versus approximately 0.253~H (MP2) and 0.221~H (CCSD), exceeding the CCSD value by more than a factor of two and highlighting amplified correlation under magnetic confinement. The force constant, $k$, (Tables S1-S3 of SI) softens from 0.403~H/Bohr$^2$ (HF) to 0.391~H/Bohr$^2$ (MP2) and 0.370~H/Bohr$^2$ (CCSD) at zero field ($\approx$ 8\% reduction) and from approximately 2.506 to 2.296~H/Bohr$^2$ at $B=3.0$~a.u. ($\approx$ 8.4\%), leading to vibrational frequencies decreasing from 4599.7~cm$^{-1}$ (HF) to 4527.5~cm$^{-1}$ (MP2) and 4405.8~cm$^{-1}$ (CCSD), i.e., a 4.4\% HF overestimate. Anharmonicity, $\omega_e\chi_e$, is especially sensitive (Tables of S1-S3 of SI), rising at zero field from 44.0~cm$^{-1}$ (HF) to 51.0~cm$^{-1}$ (MP2) and 65.2~cm$^{-1}$ (CCSD), a 48\% increase, and at $B=2.0$~a.u. from 93.1~cm$^{-1}$ (HF) to 206.8~cm$^{-1}$ (CCSD), implying a more than 55\% HF underestimation. The well width $\Delta r_{\mathrm{FWHM}}$ contracts from 2.055~Bohr (HF) to 1.908~Bohr (MP2) and 1.688~Bohr (CCSD) at zero field (about an 18\% reduction), while the flattening index increases from 1.416 to 1.656 and 2.149 (over a 50\% enhancement), confirming increased asymmetry and outer-wall softening. Together with compression and stiffness ratios (all quantities reported in Tables S1-S3 of SI), these trends show that HF exaggerates magnetic hardening, relative to CCSD, the latter being exact for $\mathrm{H}_2$ in a given finite basis set. CCSD thus provides the most balanced and physically consistent description over the full field range. For data of PEC and its characteristics for all the electron-correlation methods see supplementary Tables S1-S4.

\begin{figure}[htbp]
    \centering
    \begin{minipage}{0.98\linewidth}
        \centering
        \includegraphics[width=\linewidth]{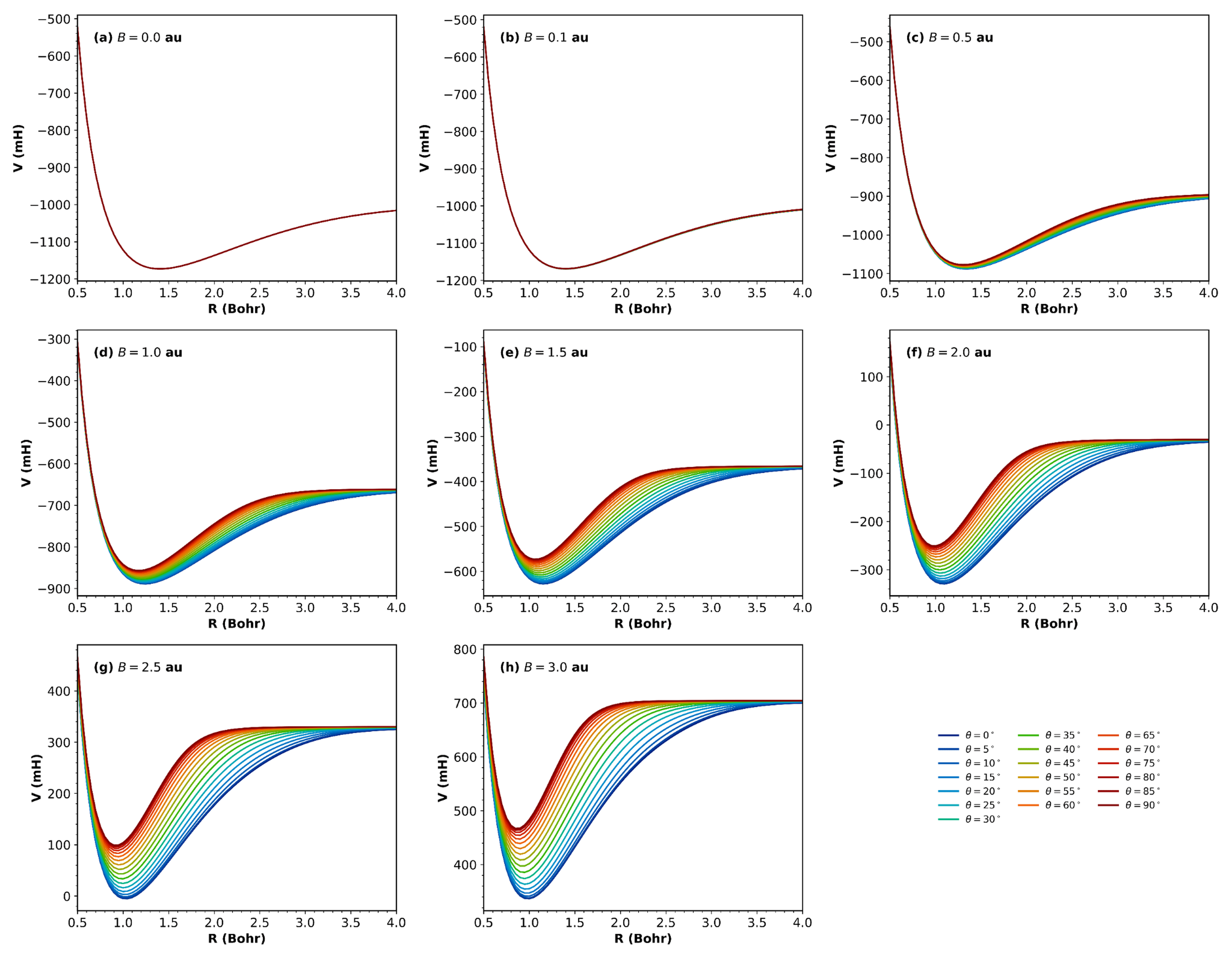}
    \end{minipage}
    
\caption{PES of $^1H_2$ computed at the CCSD/Lu-aug-cc-pVTZ level under various magnetic field strengths at various orientations:
(a) 0 au
(b) 0.1 au
(c) 0.5 au
(d) 1 au  
(e) 1.5 au  
(f) 2 au
(g) 2.5 au
(h) 3 au}

    \label{fig:H2 PES-3D}
\end{figure}

\begin{figure}[htbp]
    \centering
    \begin{minipage}{0.98\linewidth}
        \centering
        \includegraphics[width=\linewidth]{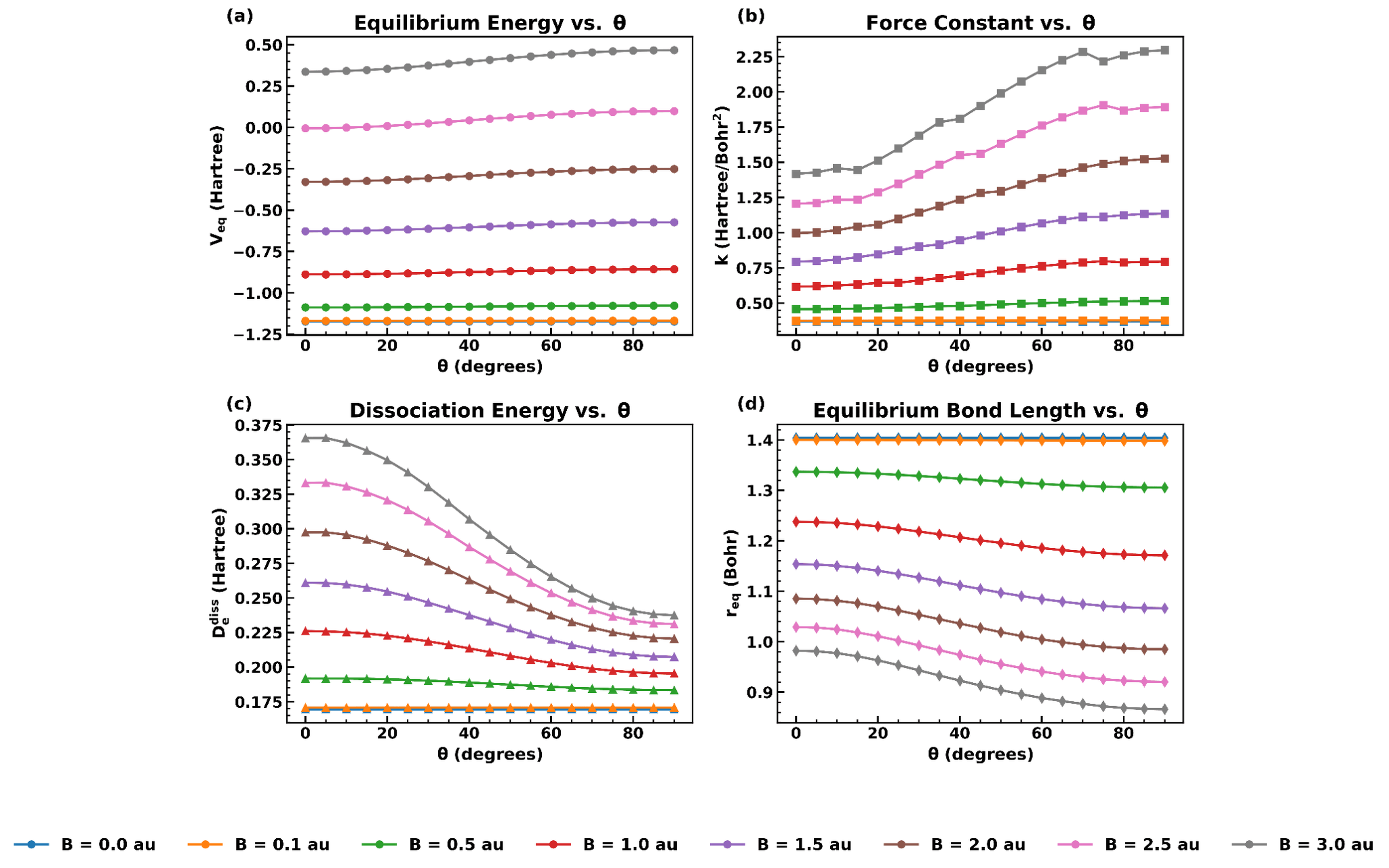}
    \end{minipage}
    
\caption{PES fitting parameters as a function of $\theta$ and $\mathrm{B}$}

    \label{fig:H2 PES-3D1}
\end{figure}

\begin{figure}[htbp]
    \centering
    \begin{minipage}{0.98\linewidth}
        \centering
        \includegraphics[width=\linewidth]{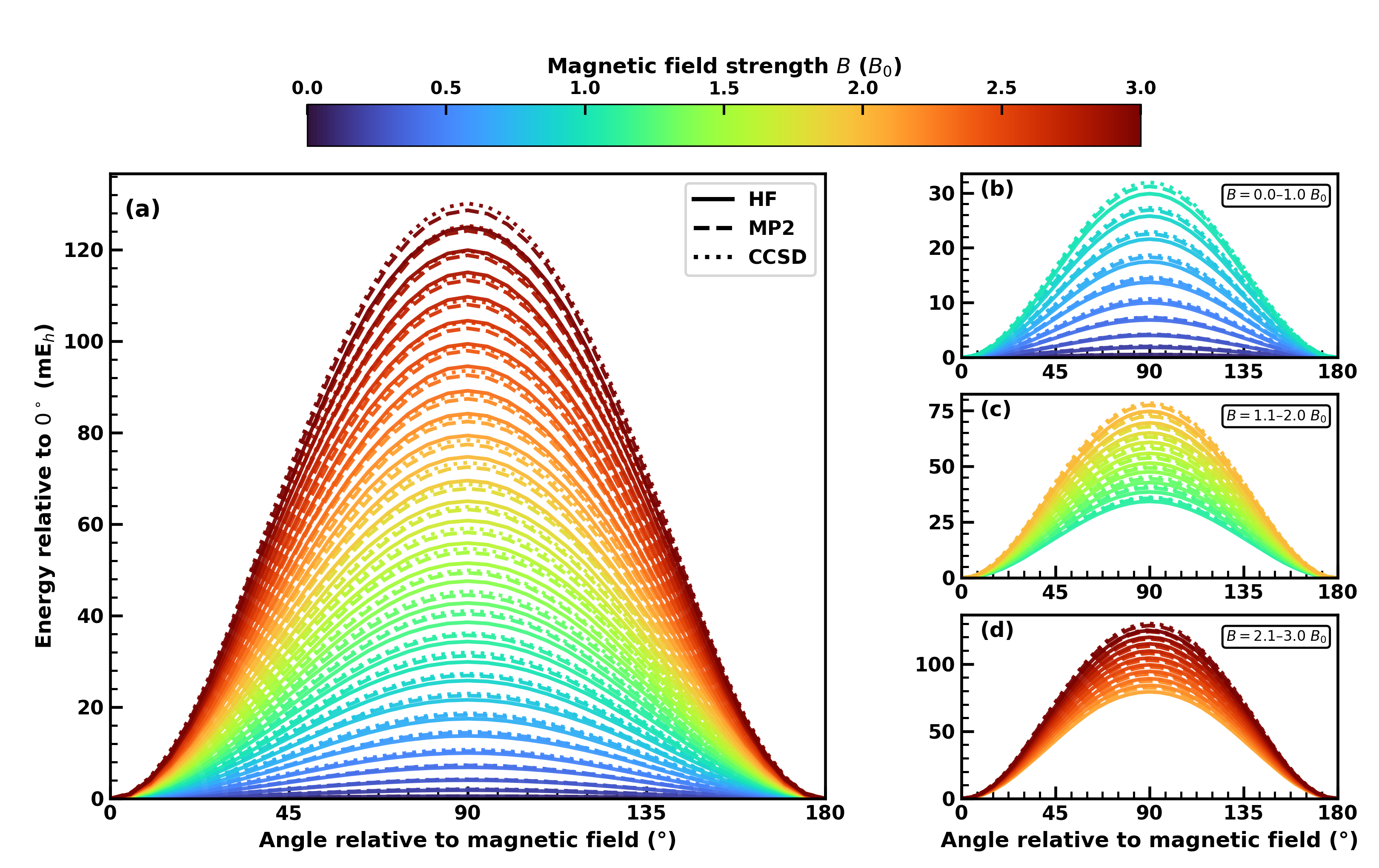}
    \end{minipage}

\caption{Angular dependence of the equilibrium energy shift
$\Delta V_{\mathrm{eq}}$ of H$_2$ in an external magnetic field.
Panel (a) shows the global anisotropy across all magnetic field strengths,
while panels (b), (c) and (d) resolve the behavior in the low-, intermediate-,
and high-field regimes, respectively.}

    \label{fig:H2 PES-anisotropy}
\end{figure}

\textbf{3D PES}: In Fig.~\ref{fig:H2 PES-3D} each subplot is the PEC at a given field while varying $\theta$ (angle between molecular axis and magnetic field) from $0^{o} \; \text{to} \;90^{o}$, where the blue end of the spectrum corresponds to lower angles and the red end of the spectrum corresponds to higher angles. The complete data from fitting and characterization of the $3D$ PES is provided in SI Tables $\mathrm{S5-S8}$. It is noticeable that as the field varies, the orientation preference of $^1\mathrm{H}_2$ at all bond lengths increases, ie. the extent of anisotropy with respect to orientation between $0^{o}$ and $90^{o}$ increases. In Fig.~\ref{fig:H2 PES-3D1} the characteristics of the PES with respect to orientation - equilibrium energy ($\mathrm{V_e}$), force constant ($k$), dissociation energy ($\mathrm{D_e}$), and equilibrium bond length ($\mathrm{R_e}$) are shown.

At B = 0, the PES is strictly isotropic with respect to molecular orientation, and $V_{eq}$ is independent of $\theta$. It can be seen from Fig.~\ref{fig:H2 PES-anisotropy}a that for all magnetic-field strengths considered, the parallel configuration ($\theta = 0^\circ$) is preferred over the perpendicular configuration ($\theta = 90^\circ$). The dissociation regime is insensitive to orientation and consequently $\mathrm{D_e}$ shows a decreasing trend in Fig.~\ref{fig:H2 PES-3D1}c as $\theta$ increases. The characterization of $V_{eq}$ with respect to $\theta$ as a normalized anisotropy is discussed in Sec. $1.4.1$ of the SI and tabulated in Table S9. The force constant ($k$) can be observed in Fig.~\ref{fig:H2 PES-anisotropy}b, to increase with $\theta$ and $B$ by up to 62\% and 84\% respectively. From Fig.~\ref{fig:H2 PES-anisotropy}d it is observable that the bond length ($\mathrm{R_e}$) decreases with respect to both field and orientation($\theta$) by up to 38.3\% and 11.8\% respectively. 

The orientation preference at the equilibrium bond length for each field strength can be quantified as the rotational energy barrier for $^1\mathrm{H}_2$ moving from parallel ($0^{o}$) to anti-parallel ($180^{o}$) orientation as depicted in Fig.~\ref{fig:H2 PES-anisotropy}. For all $B$ and all levels of theory, $\Delta V_{\mathrm{eq}}$ increases with angle, vanishing at $\theta=0^\circ$ and reaching a maximum at $\theta=90^\circ$ leading to a preference of $^1\mathrm{H}_2$ to be parallel to the field. The rotational barrier increases strongly with $B$ exceeding $\sim 100$~mH, by $B=2.5$ au resulting in field induced hindered rotation. Panels Fig.~\ref{fig:H2 PES-anisotropy}b-~\ref{fig:H2 PES-anisotropy}d, which resolve low- ($B=0.0$-$1.0$ au), intermediate- ($B=1.1$-$2.0$ au), and high-field ($B=2.1$-$3.0$ au) regimes, show a progressive amplification and steepening of $\Delta V_{\mathrm{eq}}(\theta)$, culminating in a highly warped PES at strong fields that is far from separable into radial and angular coordinates. 
  
A clear and systematic hierarchy among the three electronic-structure methods can be noted in the rotational barrier. At all magnetic fields, the dotted CCSD curves show the highest  barriers, followed by the dashed MP2 curves, and then the solid HF curve, demonstrating that electron correlation \emph{enhances} the angular anisotropy of the energy surface. In the weak-field regime ($B \lesssim 1.0\,B_0$), the angular barriers are small, and the separation between HF, MP2, and CCSD remains modest, indicating that correlation plays only a minor quantitative role when magnetic interactions are perturbative. However, in the intermediate and strong-field regimes ($B \gtrsim 1.5$-$2.0\,B_0$), the barriers grow rapidly, and the differences between the three methods become pronounced. 

Consequently, HF underestimates the hinderance to rotation (while overestimating the bond strength) relative to CCSD, MP2 being intermediate for both parameters. This behavior is consistent with the physical situation that when the molecular axis is aligned with the field, electronic motion along the field direction remains relatively delocalized, whereas for perpendicular orientations cyclotron confinement becomes more effective, raising the electronic energy. From a rovibrational standpoint, this smooth angular dependence implies that rotational motion is progressively transformed into angular libration, leading to systematic splitting and shifting of rotational levels rather than sudden rotational quenching.

Together, these trends demonstrate that molecular orientation becomes an energetically active coordinate in strong fields, leading to a fundamentally three-dimensional PES topology with strong vibration-rotation coupling, field-dependent rotational constants, and field-mediated couplings and splittings, all of which require a fully three-dimensional and correlated description for quantitative accuracy. 

\subsection{Rovibrational Spectra}
For the calculation of the rovibrational spectra of $^1\mathrm{H}_2$, we have solved the time-independent nuclear Schrodinger equation using the grid-based method as given in Eq.~\ref{Discrete_H2D}. We have computed transitions among $n_{states} = 30 $ or $120$ depending on the Boltzmann population distribution at three temperatures, T = 0 K, 1500 K, 12000 K, as these temperatures correspond to the simplest situation, previous literature studies, and the representative surface temperature of magnetic white dwarf stars, respectively. In the 2D case, the $H_2$ system is constrained to move in the xy plane only with the field along the $\mathrm{z}$-axis.

To describe the nuclear wavefunction of $^1\mathrm{H}_2$ in 2D, an xy square grid with grid length L = 9.0 au,  $\mathrm{n_{grid}} = 602$ and $\Delta x = 0.015 \;\mathrm{au}$ has been considered. In 3D, a cubic grid with L = 9.0 au, $\mathrm{n_{grid}}=301$ and $\Delta \mathrm{x} = 0.015 \;\mathrm{au}$ has been considered. The transitions between pairs of levels up to n = 30 or 120 are considered, and after calculating their respective transition quadrupole moments and oscillator strengths, the transitions which have transition quadrupole moment integrals greater than $10^{-2}$ and consequently, oscillator strengths greater than 
$10^{-27}$~au are considered allowed transitions. In this way the T=0K transitions have been captured as line spectra. For T=1500K and T=12000K the line spectra have been broadened using Lorentzian broadening with $\omega\pm 20 \;\mathrm{cm}^{-1}$.
The oscillator strengths (intensities) have been computed using Eq.~\ref{eq:quadrupole_osc}  using the length gauge representation of the transition moment integral. All transition values are in $\mathrm{cm}^{-1}$ and magnetic fields in au. The units have been suppressed henceforth for brevity. 

In order to identify and segregate the effects of changing force constants and anharmonicities on the shifts and splittings of the spectral peaks and intensities, we have carried out a sequence of computations of increasing complexity. First, a set of 2D computations is presented with a) isotropic 2D harmonic and b) 2D numerical electronic potential. Second, the $^1\mathrm{H}_2$ molecule is allowed full 3D freedom of motion in conjunction with a) isotropic 3D harmonic and b) numerical electronic 3D potentials. 
The force constants ($k$) for the isotropic harmonic potentials are obtained through a Morse fitting of the corresponding PECs as discussed in Sec.~\ref{subsection:PES} and tabulated in the SI Table S1.
In each case, the magnetic field effect is studied in 3 steps: a) Magnetic field effect only on the nucleus (denoted as Only Nucleus) - electronic Schrodinger equation is field-free, only the nuclear kinetic energy contains the field b) Magnetic field effect on only electrons (denoted as Only Electron) - electronic Schrodinger equation is field-dependent, the nuclear kinetic energy does not contain the field. c) Magnetic field effect on both nucleus and electron (denoted as Nucleus+Electron) - both electronic and nuclear Schrodinger equations contain the field. The force constants for the harmonic potentials used in these three cases are obtained as follows: a) Only Nucleus: k of B=0 electronic PEC ($E_{el}(0)$), b) Only Electron: k(B) from $E_{el}(B)$, and c) Nucleus+Electron: k(B) from $E_{el}(B)$.

\textbf{Validation of zero field spectra:}
Before going into a detailed discussion of field effects, first let us recap the rovibrational spectra of $H_2$ at zero field. Consider the Dunham expansion~\cite{Dunham1, Dunham2} with the second-order approximation where  the rovibrational energy levels of the hydrogen molecule in its ground electronic state are given by
\[
E(v,J)=
\omega_e\!\left(v+\tfrac12\right)
-\omega_e x_e\!\left(v+\tfrac12\right)^2
+ B_v\,J(J+1)
- D_v\,[J(J+1)]^2+\cdots
\]
The vibrational dependence of the rotational constant is,
\[
B_v = B_e - \alpha_e\!\left(v+\tfrac12\right)+\cdots.
\]

where $\omega_e$ is harmonic vibrational frequency, $\omega_e x_e$ is vibrational anharmonicity, $B_e$ is equilibrium rotational constant, $\alpha_e$ is vibration-rotation interaction and $D_e$ is centrifugal distortion constant. For $^1\mathrm{H}_2$, the spectroscopic constants are \(\omega_e \approx 4401.21~\mathrm{cm^{-1}}\),
\(\omega_e x_e \approx 121.34~\mathrm{cm^{-1}}\), \(B_e \approx 60.853~\mathrm{cm^{-1}}\), \(\alpha_e \approx 3.0622~\mathrm{cm^{-1}}\) and \(D_e \approx 0.047~\mathrm{cm^{-1}}\), so that the numerical form becomes,
\[
E(v,J)\approx
4401.21\!\left(v+\tfrac12\right)
-121.34\!\left(v+\tfrac12\right)^2
+\big[60.853-3.0622\!\left(v+\tfrac12\right)\big]J(J+1)
-0.047\,[J(J+1)]^2+\cdots.
\]

A $B=0$, the observed transition energy values from our numerical computations are as follows : a) HO - 2D : 4405, 4405 (doubly degenerate) b)HO - 3D: 4402, 4402 and 4402 (triply degenerate) c) Electronic PEC - 2D : 236, 4157, 4381 d) Electronic PES - 3D: 352, 4140 and 4473. 

\textbf{(a) 2D and (b) 3D HO:} It is clear that the harmonic potential yields a peak value corresponding to the fundamental vibrational frequency ($\omega_e$= $4401\pm5 $) with corresponding double and triple degeneracy for 2D and 3D respectively. This provides an initial validation of our choices of grid parameters and numerical robustness.

\textbf{(c) Electronic PEC 2D:} The allowed transitions at 236, 4157 and 4381 correspond to the $S_0(0), Q_1(2) \; \text{and} \; S_1(0)$ respectively. We know that for a 2D rigid rotor, the energy is given by  $E=m^2B_0$ instead of $E=J(J+1)B_0$ (which corresponds to 3D); consequently, the quadrupolar rotational peak corresponds to $S_0(0) \; \text{with}\;\Delta m= 2 \rightarrow \Delta E = 4B_0 = 4 \;\text{x} \;59 = 236$. The 4157 peak corresponds to the fundamental vibrational transition with anharmonic correction ($\approx$ 4161=4401-2*120), and 4381 corresponds to the rovibrational coupling peak ($\approx$ 4393=4157+236).  

\textbf{(d) Electronic PES 3D:} The quadrupole allowed transitions at 352, 4140 and 4473 correspond to  $S_0(0), Q_1(2) \; \text{and} \; S_1(0)$ respectively, matching the corresponding experimental values of 352 (Ref.~\citenum{spectraN1,spectraN2,spectraN3}); 4143, 4497 (Ref.~\citenum{spectraN4, spectraN5}) closely. The difference between our computed and experimental values are in the range of $2-25 \; \mathrm{cm}^{-1}$. The fundamental harmonic vibrational frequency of 4401 should appear as 4161 (4401 - 2 X 120) after anharmonicity correction but a slight quadrupolar coupling from the rotational component further shifts it to 4143 (4161 - 2$(B_1-B_0)$). The 4473 peak corresponds to a strongly rovibrationaly coupled transition with $\Delta \mathrm{v}=1 \; \text{and} \; \Delta J = -2$ (4140+352-2$(B_1-B_0)= 4473$). The AIMD values from Ref.~\citenum{AIMD2} are 325, 4200, and 4900 and from [Ref.~\citenum{MDLM-1}], they are 680, 3760 and 5170. Ref.~\citenum{MDLM-1} seems to have picked up a rotational overtone but missed the fundamental perhaps due to their choice of the initial velocities in the MD simulation or too short propagation time. The smaller quantitative deviations can be attributed to the use of the less accurate PES at HF and MP2 levels of theory vs our CCSD PES. The absence of quantum effects like zero point energy, quantized degeneracy, etc. in the classical MD simulation may also be partly responsible. The detailed assignments of all the rovibrational levels at zero field and their comparison with earlier work are tabulated in SI Table S19.

\begin{figure}[H]
    \centering
    \centering
    \includegraphics[width=\textwidth]{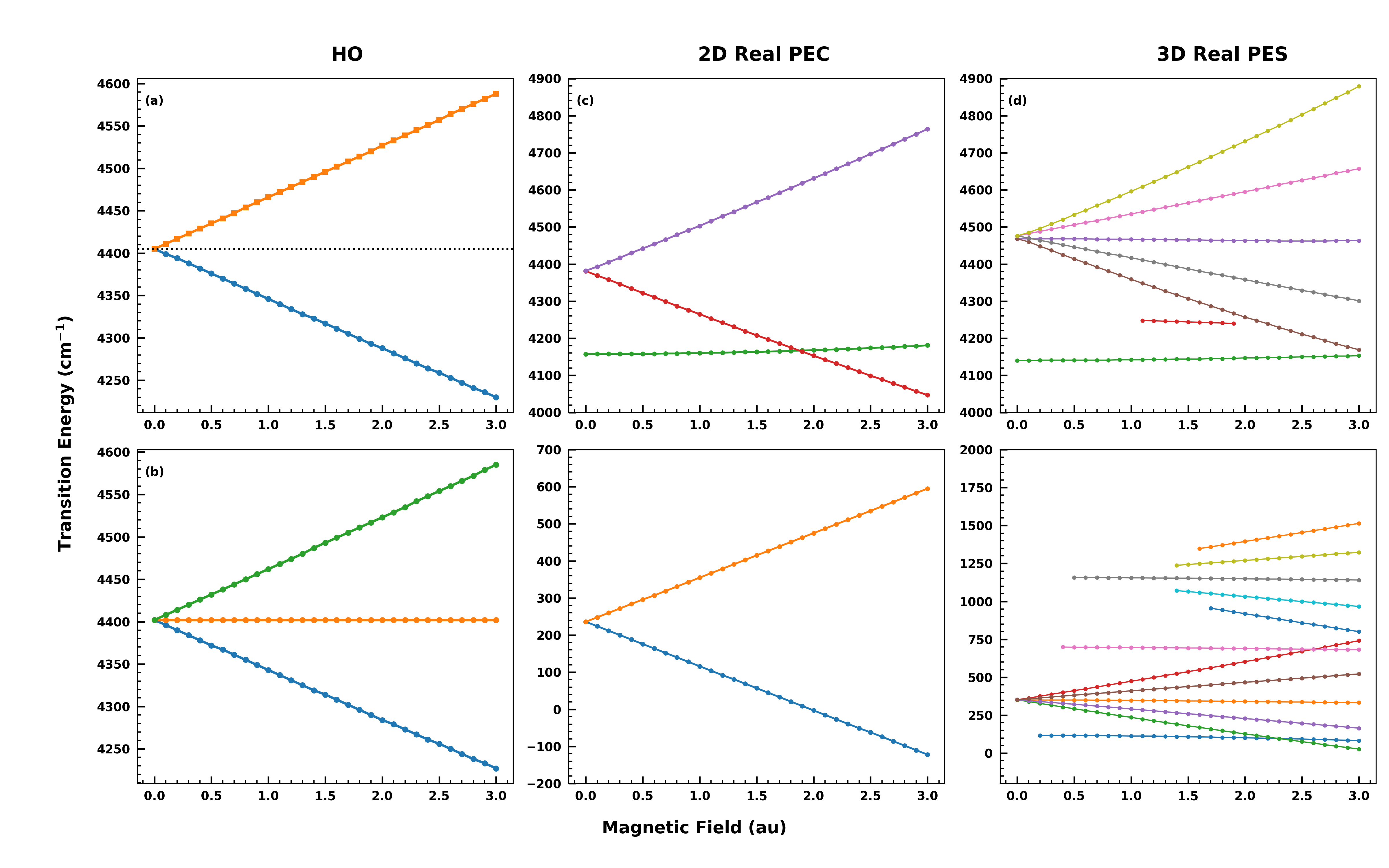}
\caption{ Transition energies vs magnetic field for the Only Nucleus case: (a) 2D HO, (b) 3D HO, (c) 2D Electronic PEC, and (d) 3D Electronic PES. (c) and (d) are split into high (top-panel) - and low (bottom panel) - energy regions. Dotted lines denote zero-field references.}
    \label{fig:all cases only nuclei}
\end{figure}

\begin{figure}[H]
    \centering
    \centering
    \includegraphics[width=\textwidth]{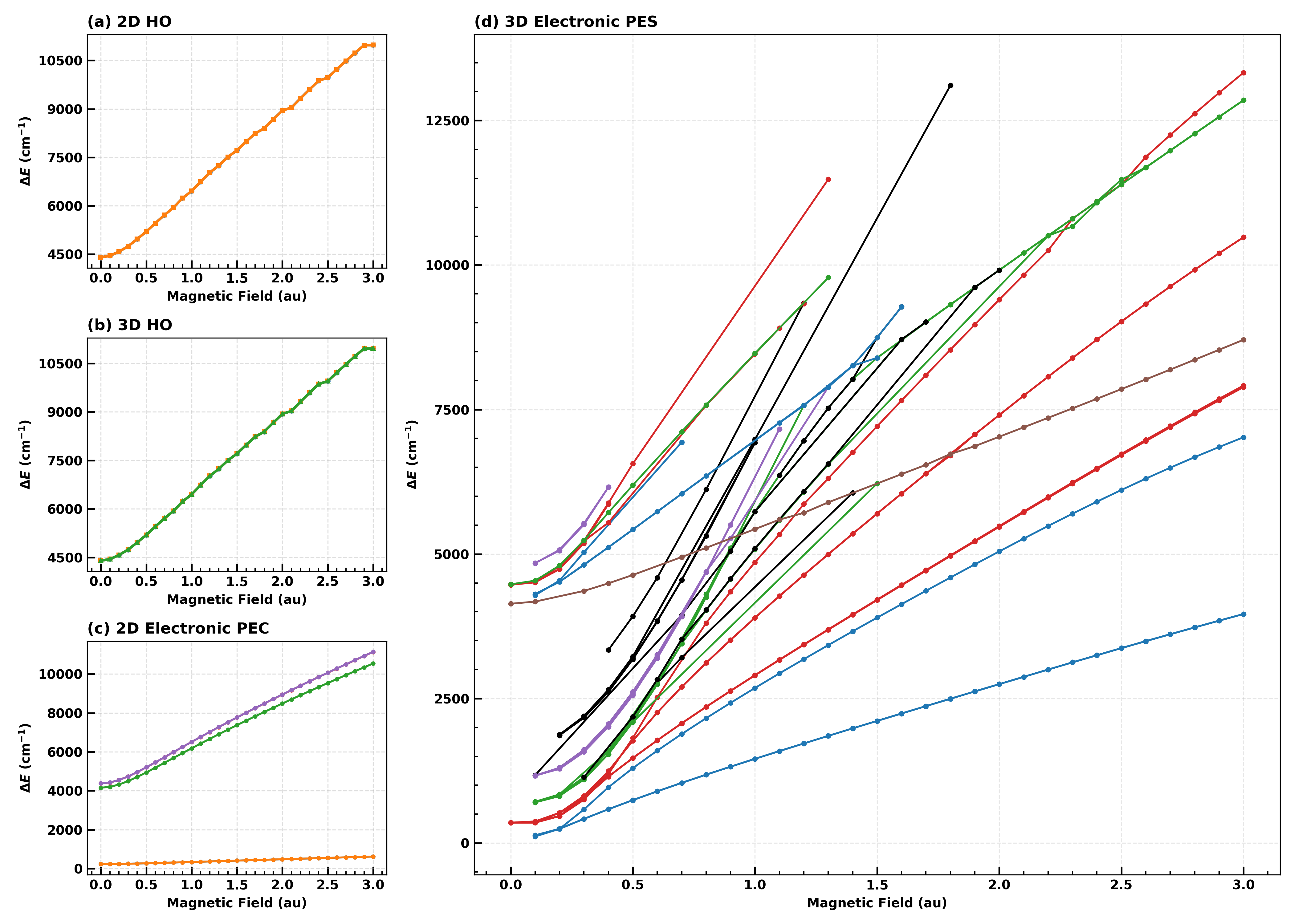}
\caption{Transition energies vs magnetic field for the only electron case: (a) 2D HO, (b) 3D HO, (c) 2D Electronic PEC, and (d) 3D Electronic PES.}
    \label{fig:all cases only electron}
\end{figure}

\textbf{Isolation of various magnetic field effects:}
The evolution of transition energies in H$_2$ under an external magnetic field reveals a complicated progression. To identify the role of the electrons and nuclei individually and thereby interpret their conjoined response to the magnetic field we can carry out exploratory computations in various artificial limiting cases. In the harmonic case, the origin of the co-ordinate system for the nuclear SE is set at the COM thereby allowing only vibrational motion. This also helps us to identify how the magnetic field affects pure vibrations. We shall describe the individual effects of the magnetic field on (1) Only Nucleus and (2) Only Electron in case of 2D and 3D with help of Fig.~\ref{fig:all cases only nuclei} and Fig.~\ref{fig:all cases only electron}. \\

\textbf{(1) Effect of magnetic field only on nucleus:}

\textbf{(a) 2D and (b) 3D HO:} The magnetic field effect on the nuclei leads to Zeeman-like splitting upto $B=10$ au (see Ref. ~\citenum{jctc1}). Thereafter, non-linear evolution leading eventually to the Landau bands can be seen. In the harmonic potential for both the 2D and 3D cases the 4401 peak splits by around $\pm$60 for every 1 au unit of increase in the field strength. In the 3D case, since energy levels are triply degenerate, the level that corresponds to $m_v$=0 is not affected by the magnetic field. The splitting features can be seen in Fig.~\ref{fig:all cases only nuclei}a and Fig.~\ref{fig:all cases only nuclei}b respectively for fields upto 3 au.

\textbf{(c) Electronic PES 2D:} The transition peaks, both vibrational and rotational now show a linear response of around 64 $\mathrm{cm}^{-1}$ akin to the HO Zeeman splitting of 60 $\mathrm{cm}^{-1}$ for every 1 au of magnetic field applied as shown in Fig.~\ref{fig:all cases only nuclei}c upto $\mathrm{B}=3 \;\mathrm{au}$. The lowest quadrupolar transition occurs between states where the final state goes below the initial state around $\mathrm{B}=2 \;\mathrm{au}$ leading to negative transition energies in Fig.~\ref{fig:all cases only nuclei}c. At $B > 2 \;\mathrm{au}$ a slight non linearity also shows up in the transition energies. Plots of energy eigenstates vs $B$ are given in SI, Fig. S9.

\textbf{(d) Electronic PES 3D:} The fundamental pure vibrational frequency line at 4141 is weakly affected by the field though it continuously blue shifts very minutely to 4152 at 3 au. However the $S_1(0)$ peak which is five fold degenerate at $4472\pm 4$ shows Zeeman splitting as expected with 2 transitions red shifting and 2 transitions blue shifting linearly with respect to $B$, though with different rates : -100, -60, 0, 60 and 134 $\mathrm{cm}^{-1}$ as shown in Fig.~\ref{fig:all cases only nuclei}e.

Likewise, the quadrupole rotational peak at 352 $(S_0(0))$ has fivefold degeneracy at $B=0$ corresponding approximately to the $m_v = -2,-1,0,1,2$ levels respectively. Its degeneracy lifting follows the same pattern as that of the $S_1(0)$ peak. However, additional interesting physics happens in the rotational regime. The dipolar rotational transitions at 118 ($\Delta J =1$), are forbidden at $\mathrm{B}=0$ but become allowed for $B\ge 0.2 \; \mathrm{au}$. Similarly, the zero-field 705.5 ($\Delta J = 3$) transition gets allowed as a peak at 699 for $B\ge 0.4 \;\mathrm{au}$. 

The $m_J =0$ component of the higher rotational peak $S_1(2)$ at 1157 gets allowed for all $B>0.4$ au, while its corresponding $m_J=\pm1$ and $m_J=\pm2$ transitions get allowed from $B=1.4$ and 1.7 respectively with Zeeman splitting at an average rate of $-50,70 \;(m_J=-1, +1)$ and $\pm120$ $\mathrm{cm}^{-1}$ ($m_J= \pm 2$) as shown in Fig.~\ref{fig:all cases only nuclei}f.

\textbf{(2) Effect of magnetic field only on Electrons (Field-dependent force constant):}

\textbf{(a) 2D and (b) 3D HO:} Within the isotropic harmonic potential, the application of a magnetic field leads to a pronounced blue shift of the transition energies. The trend of the transition energy with respect to $B$ for 2D and 3D harmonic oscillator are shown in Fig.~\ref{fig:all cases only electron}a and Fig.~\ref{fig:all cases only electron}b respectively. The kinks in the plots correspond to changes in the electronic ground state leading to a non-monotonic change in the force constant $k$ although the same state is followed, and hence, the fundamental frequency, $\omega_e$.
Quantitatively, the single vibrational peak increases from $\approx 4400$ at $B = 0$ to $\approx 11000$ at $B = 3 \; \mathrm{au}$, corresponding to a net increase of $\sim 6500$, or approximately $147\%$. The evolution is distinctly nonlinear: the increase from $B=0$ to $1$ is $\sim 2050$, from $1$ to $2$ is $\sim 2490$, and from $2$ to $3$ is $\sim 2020$. $k(B)$ is provided in SI Table S1. The rotational degrees of freedom have been disallowed in this set of calculations.

The near-perfect agreement between the 2D and 3D HO results demonstrates the stability of our numerical procedure. For instance, the transition energy at $B=0$ differs only by $\sim 3$ $\mathrm{cm}^{-1}$ (4405 vs.\ 4402), and even at $B=3$ the difference remains $\sim 15$ $\mathrm{cm}^{-1}$ (10976 vs.\ 10961), corresponding to relative deviations below $0.0015\%$ on an energy scale of $\sim 10^{4} \;\mathrm{cm}^{-1}$. This level of agreement demonstrates that, for a large range of energy and field our computational protocol is reliable. 

The degeneracies of the nuclear energy levels are independent of the field orientation, and are not lifted in this case, as the magnetic field acts exclusively on the electrons. Given the small electronic mass and larger electronic polarizability, the diamagnetic term dominates the response, leading to a uniform upward shift of the electronic potential. The harmonic model thus provides a clean reference limit in which the magnetic field induces a coherent and uniform blue shift without altering the underlying structure of the spectrum.

\textbf{(c) Electronic PEC 2D:} For B=0, as already discussed, the peak values are 236, 4157 and 4381 which correspond to $S_0(0), Q_1(2) \; \text{and} \; S_1(0)$ respectively. The rotational bands and vibrational bands evolve at different rates and their evolution with respect to $B$ can be noticed in Fig.~\ref{fig:all cases only electron}c. The lowest transition increases from $236$ at $B = 0$ to $621$ at $B = 3$, corresponding to a net shift of $\sim 385$ or $\sim 163\%$. The higher transitions evolve from $4157$ to $\approx 10537$ ($\sim 153\%$), while another increases from $4381$ to $\approx 11131$ ($\sim 154\%$).  The field dependence remains qualitatively similar to the HO approximation with a pronounced non-linear character. However, in this case, the kinks are absent as a global spline-fitted function has been used rather than just the force constant in the HO case. For the lowest transition, the increase from $B=0$ to $1$ is $\sim 105$, from $1$ to $2$ is $\sim 141$, and from $2$ to $3$ is $\sim 139$, indicating a gradual enhancement of the effective confinement. Unlike the harmonic case, however, the absolute energy scale is reduced reflecting the anharmonic nature of the molecular potential.

An important validation of our numerical protocol is the persistence of near-degenerate pairs of transitions, such as $236$-$236$ at low field and $4381$-$4381$ at higher energies. The splitting between such pairs remains extremely small (typically $\leq 1$-$2$ $\mathrm{cm}^{-1}$) across the entire field range. This is as expected as the system retains circular symmetry in the xy plane about the origin (0,0) with the reduced mass $\mu_M$ placed at $\mathbf{R}$ on the xy grid in all 4 quadrants and $B=B_z$. $E_{el}(R)$ is replicated for all directions $\theta$ in the xy-plane for the 2D case.

\begin{table}[H]
\centering
\small
\caption{3D only electron transition energy features summarized. Symbols meaning as follows: a) R means Rotational b) V means vibrational c) RV means Rovibrational d) X means higher states. An error bar of $\pm20\;\mathrm{cm}^{-1}$ has been considered for assigning degeneracy.}
\label{tab:3D-only-electron-features-summarized}

\resizebox{\textwidth}{!}{%
\begin{tabular}{|c|c|
>{\raggedright\arraybackslash}p{2.6cm}|
>{\raggedright\arraybackslash}p{2.6cm}|
>{\raggedright\arraybackslash}p{2.8cm}|
>{\raggedright\arraybackslash}p{2.3cm}|
>{\raggedright\arraybackslash}p{2.6cm}|
>{\raggedright\arraybackslash}p{2.4cm}|}
\hline
\textbf{S.No} & \textbf{Group ID} & \textbf{Low-field Energy (cm$^{-1}$)} & \textbf{Transition Set} & \textbf{Physical Type} & \textbf{Degeneracy} & \textbf{$\Delta E$ at $B=3 \; \mathrm{au}$ (cm$^{-1}$)} & \textbf{Splitting Onset (a.u.)} \\
\hline

1 & R1 & $\sim 0$ & $1 \rightarrow 2,3,4$ & Pure rotational ($J=0 \rightarrow 1$) & 3-fold & $\sim 3960$-$7020$ & $\sim 0.4$ \\

2 & R2 & 351-352 & $1 \rightarrow 5$-$9$ & Rotational ($J \approx 1$-$2$) & 5-fold & $\sim 7900$-$10480$ & $\sim 0.3$ \\

3 & R3 & $\sim 700$-$720$ & $1 \rightarrow 10$-$14$ & Rotational (higher $J$) & $\sim$5-fold & $\sim 9000$-$10000$ & $\sim 0.4$ \\

4 & R4 & $\sim 1150$-$1200$ & $1 \rightarrow 15$-$20$ & Rotational ($J \approx 3$-$4$) & $\sim$5-fold & $\sim 10000$-$11000$ & $\sim 0.5$ \\

5 & R5 & $\sim 1800$-$1900$ & $1 \rightarrow 21$-$30$ & Rotational dense manifold & $\sim$8-10 fold & $\sim 11000$-$12000$ & $\sim 0.5$ \\

6 & R6 & $\sim 2600$-$2700$ & $1 \rightarrow 31$-$40$ & High-$J$ rotational & $\sim$8-10 fold& $\sim 12000$-$12500$ & $\sim 0.6$ \\

7 & V1 & $\sim 4140$ & $1 \rightarrow 41,42$ & Fundamental vibrational ($v=0 \rightarrow 1$) & 2-fold & $\sim 11000$-$12000$ & $\sim 0.4$ \\

8 & RV1 & 4468-4476 & $1 \rightarrow 43$-$47$ & Rovibrational ($v=1$ + rotation) & 5-fold & $\sim 12800$-$13300$ & $\sim 0.4$ \\

9 & RV2 & $\sim 4500$-$5000$ & higher states & Rovibrational higher $J$ & $\sim$5-8 fold& $\sim 13000$ & $\sim 0.5$ \\

10 & RV3 & $\sim 6000$-$8000$ & upper states & Highly excited rovibrational &  & $\sim 12000$-$13500$ & $\sim 0.5$ \\

11 & X1 & $\sim 1500$-$3000$ & selected states & Mixed (non-adiabatic) & 1-2 fold & terminate early & $\sim 0.5$ \\

\hline

\textbf{Group ID} & \textbf{Nonlinear Region (B in au)} & \textbf{Linear Regime (B in au)} & \textbf{Avg Slope (cm$^{-1}$/au)} & \textbf{$m_J$ Structure} & \textbf{Correlation Across Regions} & \textbf{Key Physical Insight} & \textbf{Librational Beyond Field} \\
\hline

R1 & 0-0.6 & $>$0.6 & $\sim 1300$-$2300$ & Weak & Mirrors the RV1 weak branch & Lowest rotational ladder, weak magnetic response & 0.2 \\

R2 & 0-0.6 & $>$0.6 & $\sim 2500$-$3300$ & Clear & Strongly correlated with RV1 & First clear Zeeman fan structure & 0.3 \\

R3 & 0-0.8 & $>$0.8 & $\sim 2500$ & Moderate & Tracks RV3 splitting & Increasing angular momentum coupling & 0.4 \\

R4 & 0-1.0 & $>$1.0 & $\sim 2600$ & Strong & Matches RV4 slope trend & Transition toward mixed regime & 0.8 \\

R5 & 0-1.2 & $>$1.2 & $\sim 2800$ & Strong & Correlated with RV cluster & Dense avoided crossings begin & 1.3 \\

R6 & 0-1.3 & $>$1.3 & $\sim 2900$ & Strong & Strong RV coupling & Strong mixing, breakdown of good $J$ & 2 \\

V1 & 0-0.9 & $>$0.9 & $\sim 2600$ & Moderate & Anchor for all RV branches & Reference vibrational energy scale & -- \\

RV1 & 0-1.0 & $>$1.0 & $\sim 3000$-$3200$ & Strong & Mirrors R2 splitting pattern & Vibrational excitation enhances Zeeman response & -- \\

RV2 & 0-1.2 & $>$1.2 & $\sim 3000+$ & Strong & Tracks R3-R4 & Coupled rotational-vibrational dynamics & -- \\

RV3 & 0-1.5 & partial & up to $\sim 4000$ & Very strong & Linked to R5-R6 & Strong non-adiabatic mixing & -- \\

X1 & 0-1.5 & weak & up to $\sim 5000$ & Irregular & No clear partner & Avoided crossings dominate & -- \\

\hline
\end{tabular}
}
\end{table}

\textbf{(d) Electronic PES 3D:} A qualitatively different behavior emerges in the full three-dimensional Electronic PES. Here, nuclear motions where the molecule changes its angle with $\mathbf{{B}}$ are allowed, i.e, nuclear velocity is allowed to have a z-component or alternatively $\Psi_{\mathbf{nuc}}$ is $\Psi_\mathbf{nuc}(\mathrm{x,y,z})$. Attention is restricted to physically significant transitions with $|{\rm Quad}| > 10^{-2}$. While a global blue shift remains evident, the evolution of individual transitions becomes highly non-uniform.

The rate of increase is strongly transition-dependent. Some transitions grow rapidly and dominate the high-energy region, while others evolve more slowly, leading to a pronounced widening of the energy distribution with an increasing magnetic field. In contrast with the 2D case, near-degenerate pairs are no longer observed; instead, the spectrum becomes increasingly dispersed, with well-separated transition manifolds. The transition energy evolution of the only electron case in 3D is shown in Fig.~\ref{fig:all cases only electron}d and summarized in Table~\ref{tab:3D-only-electron-features-summarized}. The general trends are as follows: a) The rotational bands (states whose transition energy are $<$ 4000) split into multiple branches depending on the $|m_J|$. The extent to which the curves split into branches are proportional to the value of $|m_J|$. Another interesting thing to note is that there is a one-to-one correspondence between the transition energy evolution in the rotational regime and in the rovibrational regime on account of very specific coupling of rotational and vibrational transitions.  

Let us now discuss some individual transitions which are represented in Fig.~\ref{fig:all cases only electron}d. The rotational peak at 352 $\mathrm{cm}^{-1}$ ($S_0(0)$) at $B=0$ is 5-fold degenerate. For $B<0.4 \;\mathrm{au}$, they are slightly split and rotational in character. The increasing hindrance to the free rotational character can be noticed in the blue-shifting of the peaks. For $\mathrm{B}> 0.4$, librational/oscillatory character emerges as the degenerate peaks split at $\mathrm{B}=0.4 \;\mathrm{au}$ into groups of 2, 2 and 1 peaks. Peaks with energies less than the rotational barrier (assuming no tunnelling) may be considered as libration and the others as rotation. Subsequently, the gap between them increases non-linearly upto $\mathrm{B}=1 \; \mathrm{au}$ and then linearly for $\mathrm{B} > 1 \; \mathrm{au}$. The rotational barrier at $B=0.4$ is $7.30 \;\mathrm{mH}$ with rotational peaks of energy less than the barrier acquiring librational character. 

The peak at 4140.4 $\mathrm{cm}^{-1}$, $Q_1(2)$, blue shifts non-linearly for $\mathrm{B}\le0.5 \, \mathrm{au}$ and linearly for $B > 0.5 \; \mathrm{au}$ with a slope of $\approx 1530 \;\mathrm{cm}^{-1}$/au of B. 

Peaks at an average of 4472 $\mathrm{cm}^{-1}$ ($S_1(0)$) show combined effects of $S_0(0) \; \textbf{and} \; Q_1(2)$, as they correspond to the rovibrational coupling peaks as discussed earlier. For $0.4\le\mathrm{B}\le1.3 \;\mathrm{au}$ they have characteristics of the fundamental vibration coupling with the librational peak. The libration-vibration coupling is operative only for the range $0.4\le\mathrm{B}\le0.9 \;\mathrm{au}$. 

Next we shall discuss the transitions allowed only under a magnetic field. First, the rotational peak $S_0(1)$ at 117.3 $\mathrm{cm}^{-1}$ at $\mathrm{B}=0$ is quadrupole forbidden but gets allowed($|\mathrm{Quad}|>10^{-2}$) in the presence of field for $B=0.1$. At $B=0.2$, it splits into 2,1 group of peaks, where the peaks in the group of 2 shows a linear blue-shift with a rate of 1156, $\mathrm{cm}^{-1}$/au of $B$, and the single peak shows non-linear (log-type) blue shift up to $B=1.0 \; \mathrm{au}$ and linear blue-shift $B>1.0 \; \mathrm{au}$ with a 
rate of $1690 \; \mathrm{cm^{-1}}$/au of B.

There are a few other minor rovibrational coupling peaks at $4287,4308,4841,4856,4862$ which become allowed at $B=0.1$ and linearly blue-shift up to a certain field and again become disallowed in the harmonic/Landau regime.  

The behavior of the various peaks discussed above reflects  the inherently anisotropic nature of the full 3D electronic PES. The magnetic field, acting on the electronic structure, does not simply renormalize the energy scale but modifies different regions of the electronic density in distinct ways. As a result, the effective potentials experienced by different states evolve differently, giving rise to state-dependent scaling and a clear breakdown of the uniform behavior observed in both the harmonic and 2D electronic cases.

\textbf{(3) Effect of magnetic field on both Nuclei and Electrons}
\begin{figure}[H]
    \centering
    \centering
    \includegraphics[width=\textwidth]{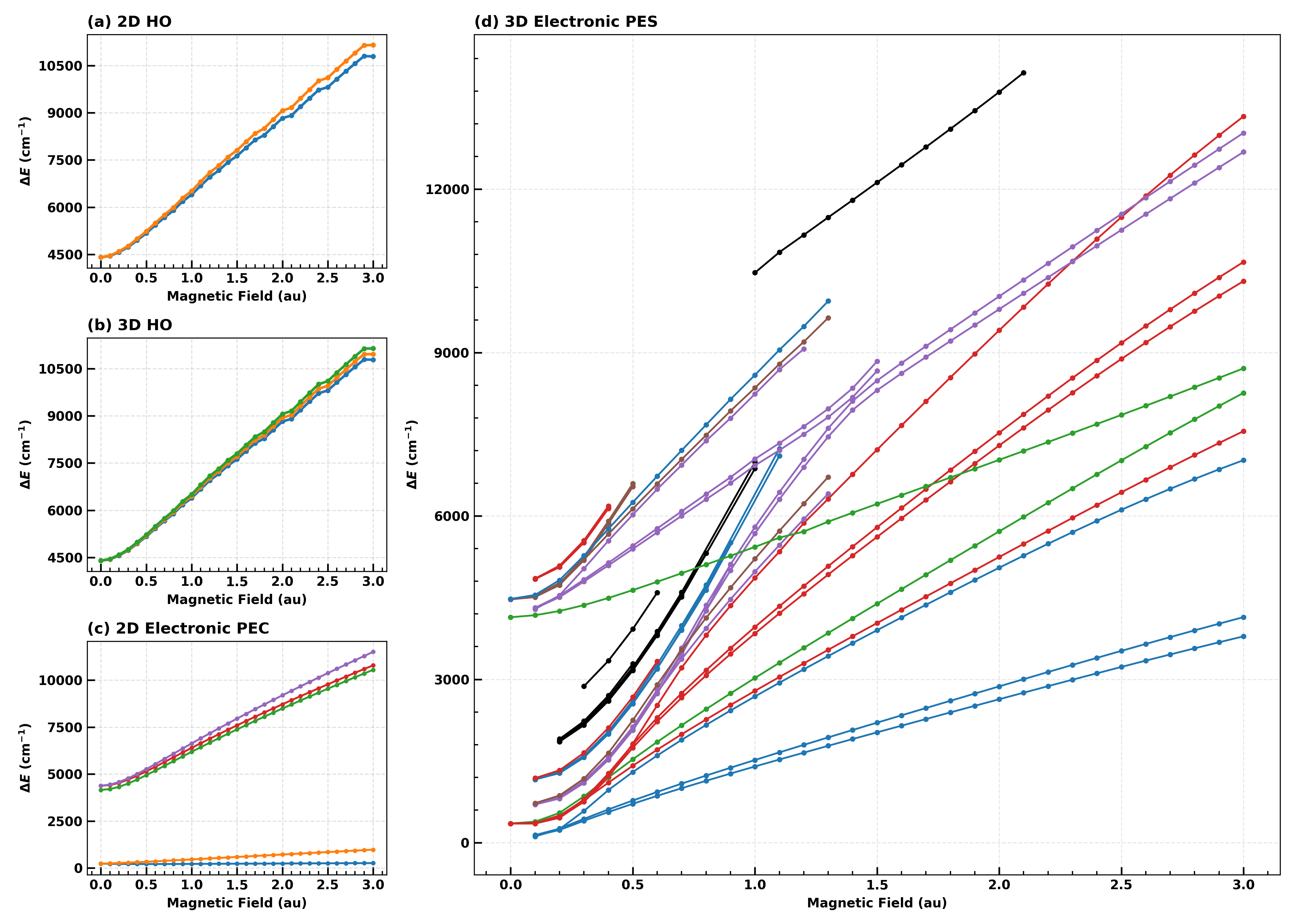}
\caption{Transition energies vs magnetic field on both nuclei and electrons: (a) 2D HO, (b) 3D HO, (c) 2D Electronic PEC, and (d) 3D Electronic PES.}
    \label{fig:all cases both nucleus and electron}
\end{figure}

\begin{figure}[H]
    \centering
    \includegraphics[width=0.7\linewidth]{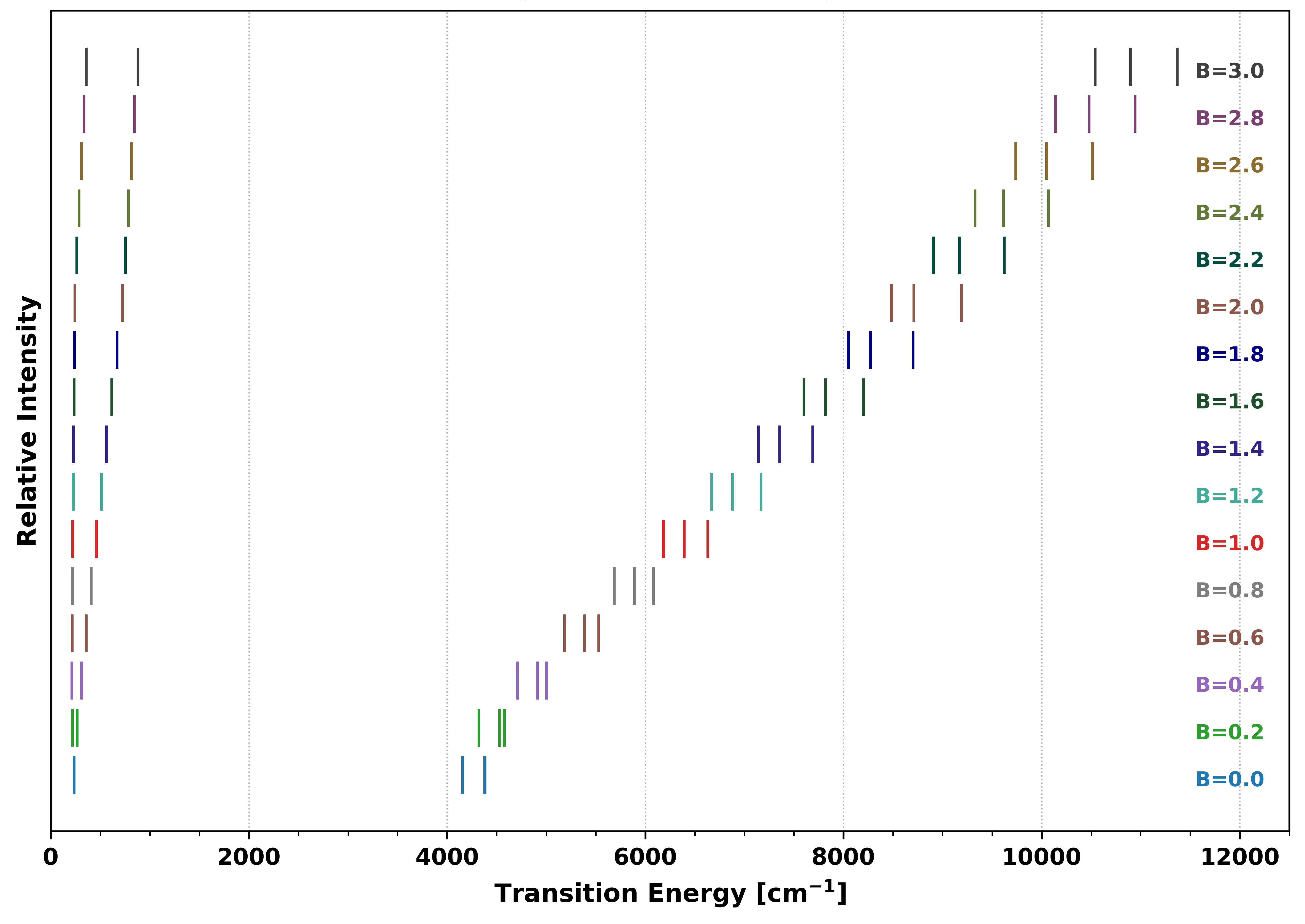}
    \caption{2D Electronic Potential (Magnetic field acting on both nuclei and electrons) : Rovibrational Line Spectra of $H_2$ subjected to perpendicular magnetic field at $T = 0$ K}
    \label{fig:2D_Num_NE_T0K}
\end{figure}

\begin{figure}[H]
    \centering
    \includegraphics[width=0.7\linewidth]{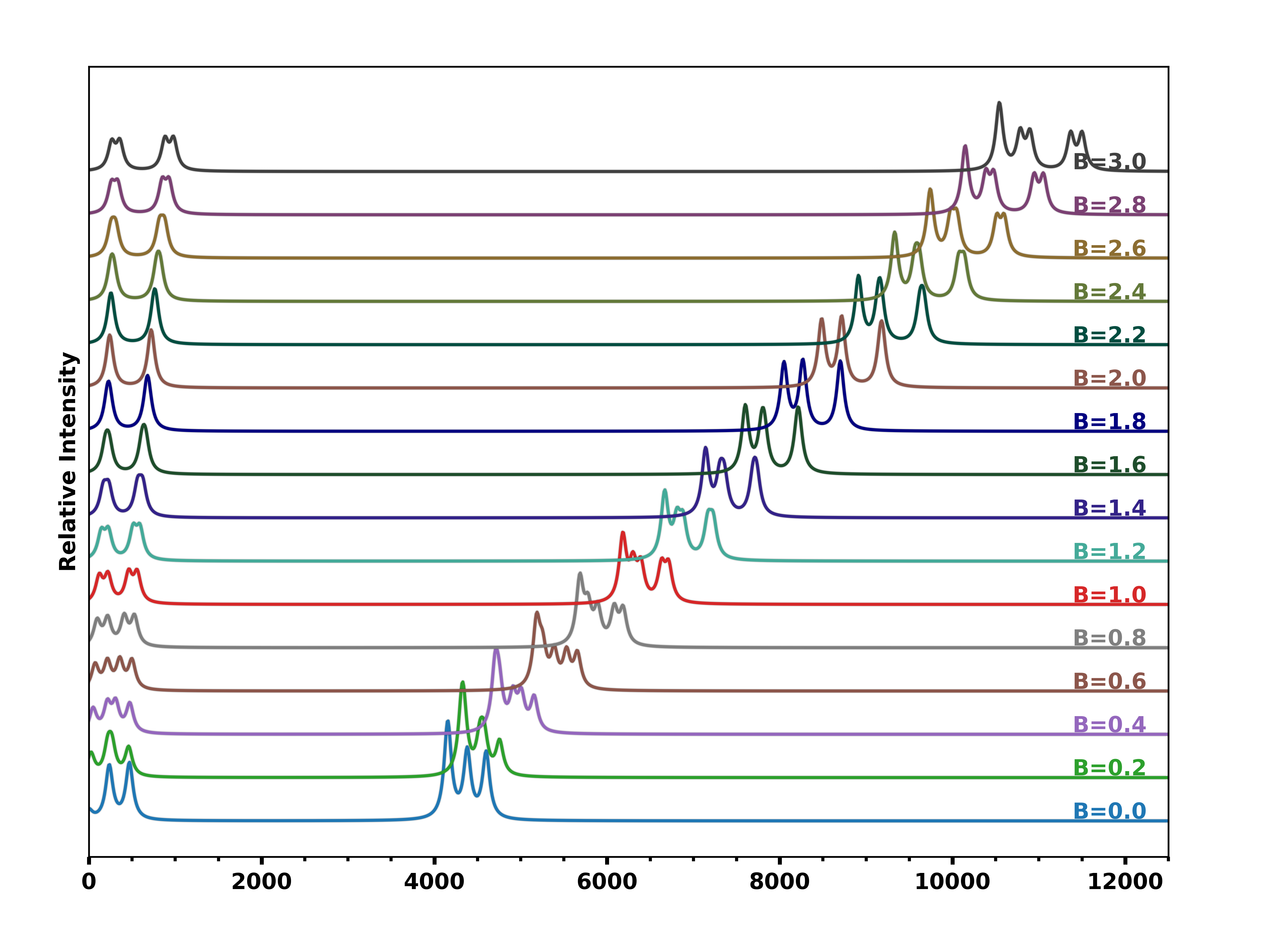}
    \caption{2D Electronic Potential (Magnetic field acting on both nuclei and electrons) : Rovibrational Line Spectra of $H_2$ subjected to perpendicular magnetic field at $T = 1500$ K}
    \label{fig:2D_Num_NE_T1500K}
\end{figure}

\begin{figure}[H]
    \centering
    \includegraphics[width=0.7\linewidth]{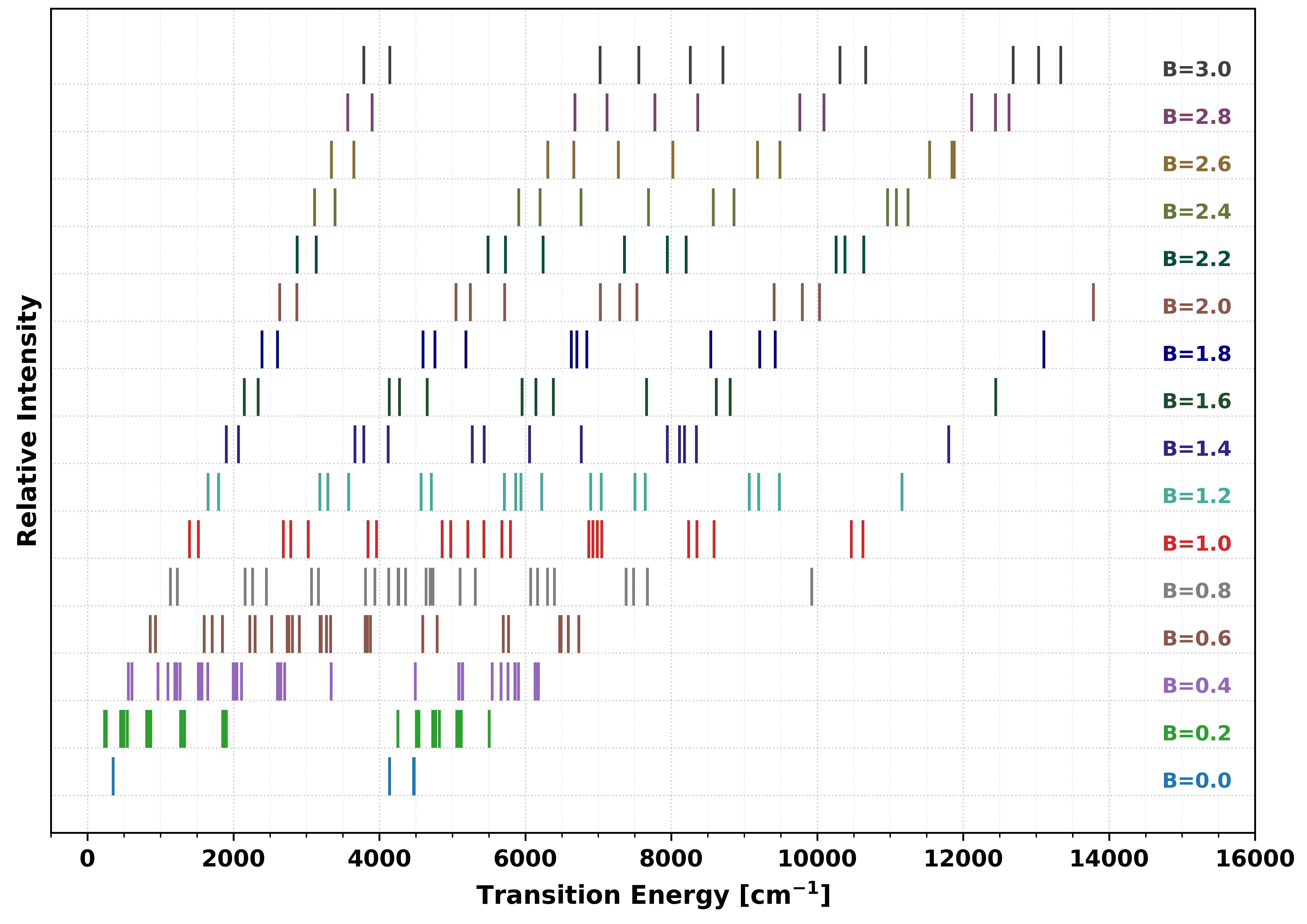}
    \caption{3D Electronic Potential (Magnetic field acting on both nuclei and electrons) : Rovibrational Line Spectra of $H_2$ subjected to magnetic field at $T = 0$ K}
    \label{fig:3D_Num_NE_T0K}
\end{figure}

The evolution of the transition energies with B can be seen in Fig.~\ref{fig:all cases both nucleus and electron}. The quadrupolar spectra of $0\mathrm{K}$ and $1500\mathrm{K}$ temperatures can be seen in Figs.~\ref{fig:2D_Num_NE_T0K} and ~\ref{fig:2D_Num_NE_T1500K} respectively for 2D and Figs.~\ref{fig:3D_Num_NE_T0K} and ~\ref{fig:3D_Num_NE_T1500K} for 3D respectively. Note that in Fig. 10 all the spectral lines with $|\mathrm{Quad}|> 10^{-2}$ (quadrupole TMI) have been drawn with equal intensity. In Fig. ~\ref{fig:2D_Num_NE_T1500K}, the actual oscillator strengths along with a Lorentzian broadening of $\Delta \nu =20\;\mathrm{cm}^{-1}$ has been considered.

\textbf{(a) 2D and (b) 3D HO:} In the HO cases the effects on the nuclei and electron are additive - that is the transition energies shift according to the pattern followed in the only electron case and the splitting of the degenerate peaks follows the only nuclei case. A splitting rate of $60 \; \mathrm{cm}^{-1}$/au of B is observed here as well. 

\textbf{(c) Electronic PEC 2D:} In the electronic 2D PES also, when the magnetic field is considered on both nuclei and electron, the effects are also additive as discussed above. 
The $S_0(0)$ rotational peak at 236 $\mathrm{cm}^{-1}$ and $S_1(0)$ rovibrational coupling peak at 4381 which is doubly degenerate at $B=0$ splits and shifts when a field is applied. They split into two distinguishable peaks with the splitting corresponding to the cyclotron frequency of $240 \;\mathrm{cm}^{-1}$/au of B applied.

The $Q_1(2)$ fundamental vibrational peak at 4157 $\mathrm{cm}^{-1}$, shows only blue shifting with $B$ exactly as in the only electron case.

The $S_1(0)$ rovibrational coupling peak at 4381 = 4157+236-2($B_1-B_0$) at $B=0$ shows interesting characteristics. It has four combined effects: a) the effects of B on the electrons and nuclei for the rovibrational $S_0(0)$ peak, which results in blue shifting and splitting; b) the effect of B on the electronic component of the fundamental vibrational $Q_1(2)$ - major cause of blue shifting; c) the effect of $B$ on the rotational constants; and d) the cyclotron coupling and splitting of the $S_1(0)$ as $S_1(0) = Q_1(2)+S_0(0)-2(B_1-B_0)$. In short, it blue shifts with contributions from hindered rotational characteristics, acquiring of $S_0(0)$ character and vibrational enhancement of $Q_1(2)$ as the equilibrium bond length decreases with an increase in $B$. Also, its splitting pattern is a result of the cyclotron coupling of $S_0(0)$ though slight deviations are present. The transition energy shows the additive effect of shifting from electronic effects as in Fig.~\ref{fig:all cases only electron}c and splitting of the respective peaks by 240 $\mathrm{cm}^{-1}$/au of increase in $B$. The resultant spectra is presented in Fig.~\ref{fig:2D_Num_NE_T0K}. 

When a temperature of T=1500K is considered, the second level is populated leading to additional transitions in Fig.~\ref{fig:2D_Num_NE_T1500K}. The characteristics of $S_1(0), S_0(0) \; \text{and} \; Q_1(2)$ follow similar patterns as at $T=0K$. In addition, further branching out of $S_0(0)$ takes place as it couples with one of its own branches with a characteristic coupling constant on top of cyclotron coupling, giving rise to two additional branches. This phenomenon can be attributed to magnetically induced self-coupling, similar to that of overtones. Consequently, the rovibrational peak $S_1(0)$ also experiences multiple rovibrational couplings leading to further splitting in addition to the split from cyclotron coupling. However, the fundamental vibration $Q_1(2)$ remains similar to the $T=0K$ situation, with the exception of becoming a doubly degenerate transition.

\begin{figure}[H]
    \centering
    \centering
    \includegraphics[width=0.92\textwidth]{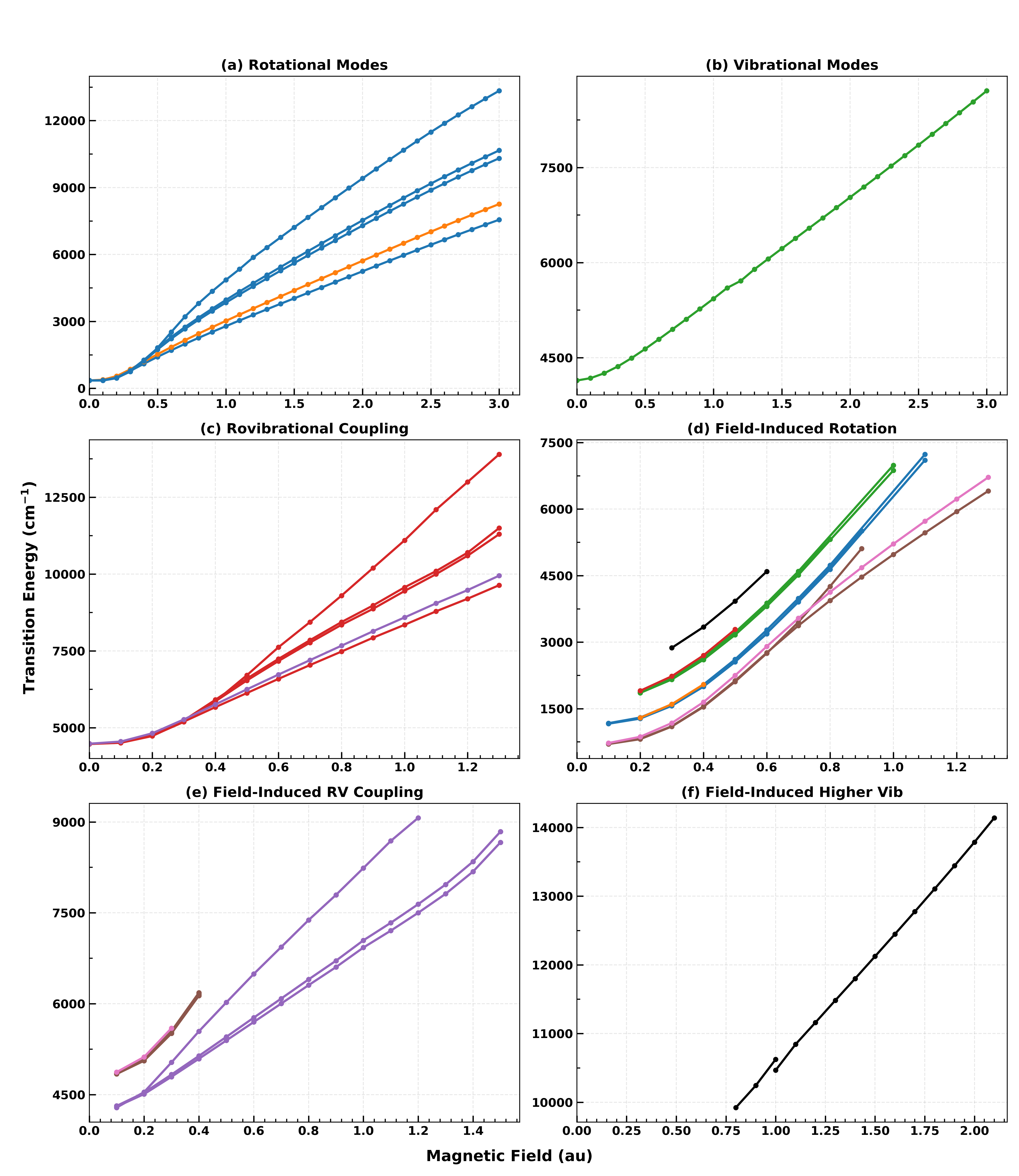}
\caption{Magnetic-field dependence of transition energies for the 3D electronic PES of $H_2$ with coupled electron–nuclear motion: (a) rotational, (b) vibrational, (c) rovibrational coupling, (d) field-induced rotational, (e) field-induced rovibrational coupling, and (f) field-induced higher vibrational modes. Transition energies are shown as a function of magnetic field (au), with color grouping indicating low-field spectral proximity.}

    \label{fig:3D-resolved-te-T=0K}
\end{figure}

\begin{figure}[H]
    \centering
    \centering
    \includegraphics[width=0.92\textwidth]{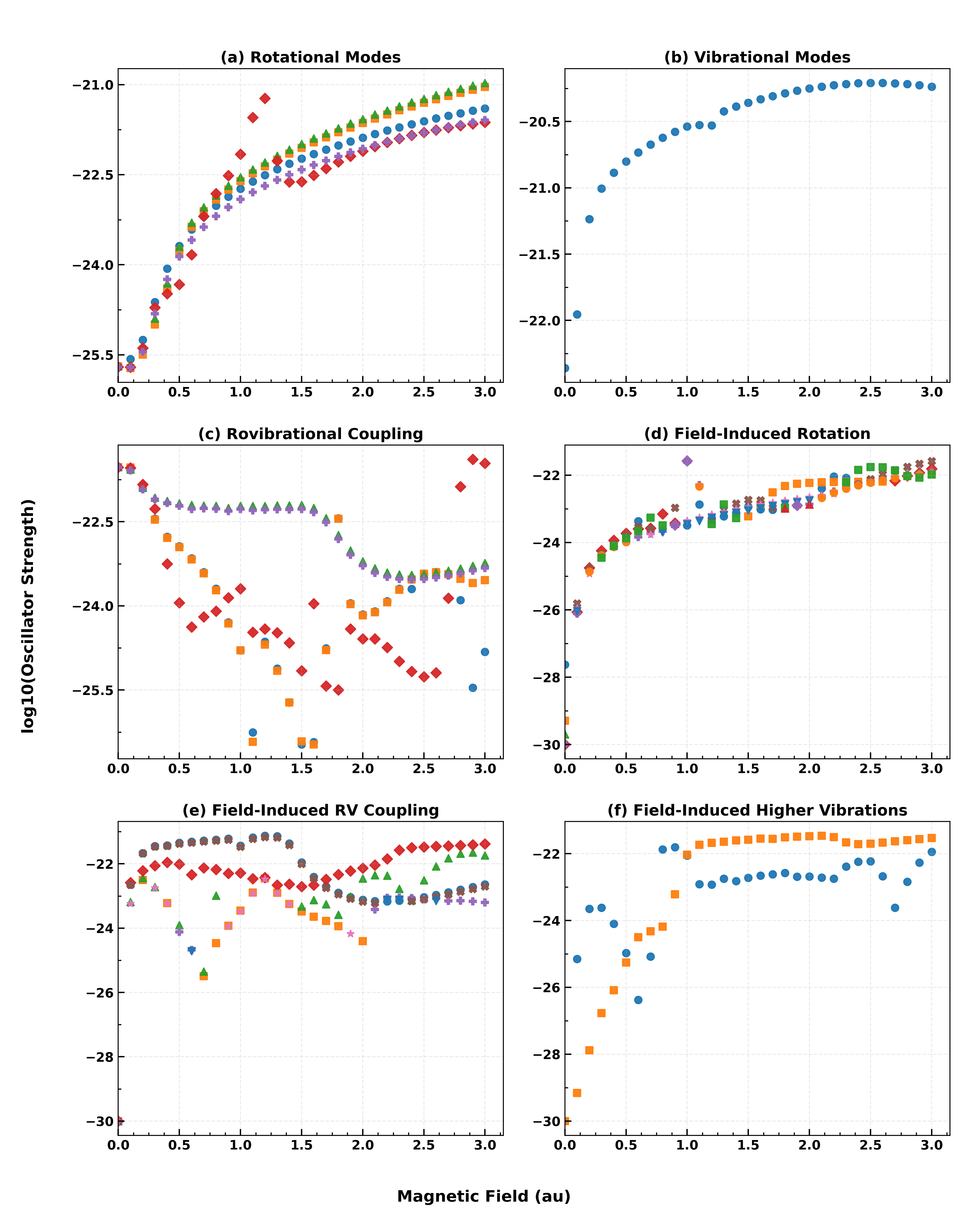}
\caption{Magnetic-field dependence of oscillator strengths for the 3D electronic PES of $H_2$ with coupled electron–nuclear motion: (a) rotational, (b) vibrational, (c) rovibrational coupling, (d) field-induced rotational, (e) field-induced rovibrational coupling, and (f) field-induced higher vibrational modes. Transition energies are shown as a function of magnetic field (au).}

    \label{fig:3D-resolved-os-T=0K}
\end{figure}

\begin{figure}[H]
    \centering
    \includegraphics[width=0.7\linewidth]{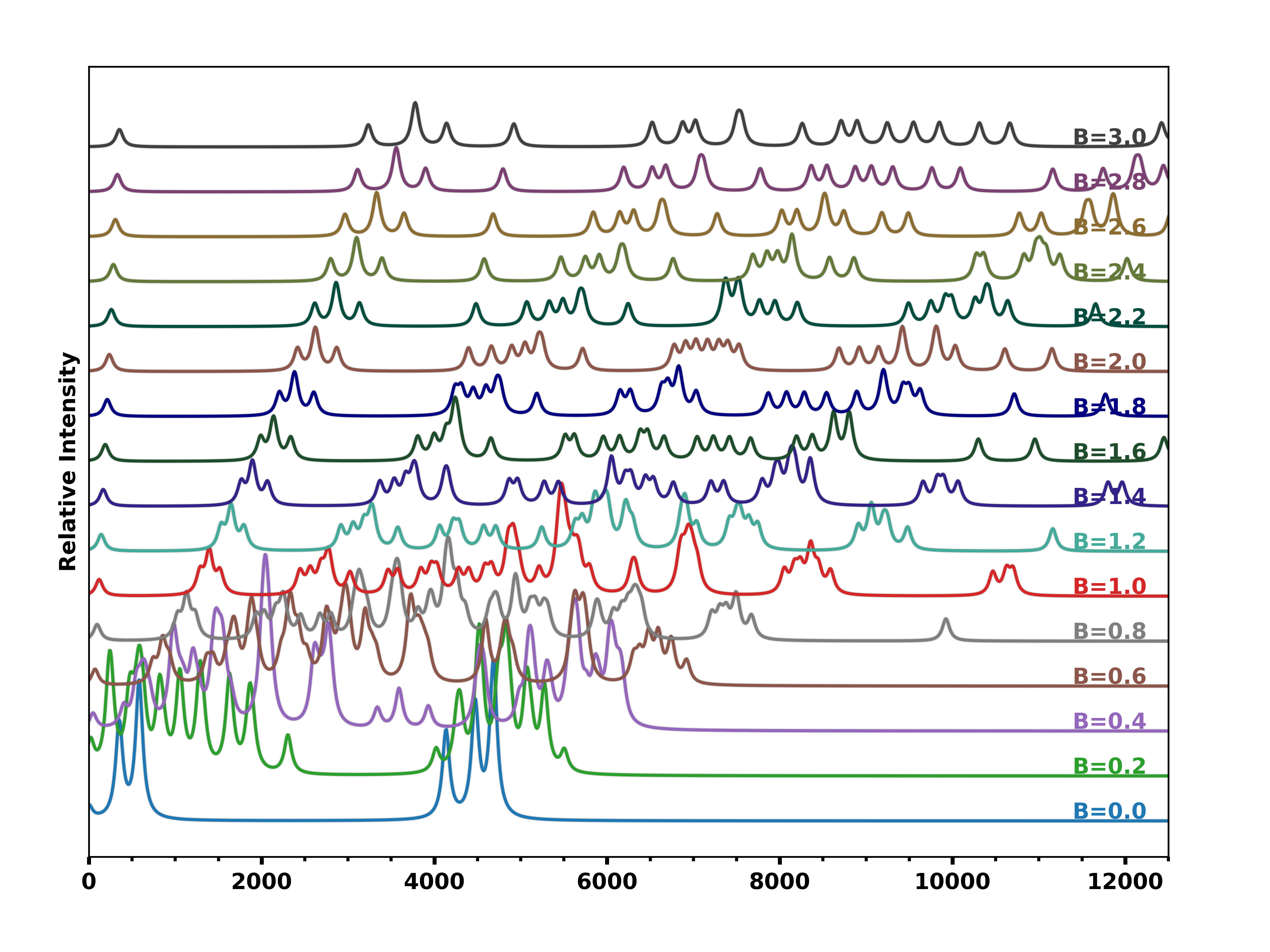}
    \caption{3D Electronic Potential (Magnetic field acting on both nuclei and electrons) : Rovibrational Line Spectra of $H_2$ subjected to magnetic field at $T = 1500$ K}
    \label{fig:3D_Num_NE_T1500K}
\end{figure}

\textbf{(d) Electronic PES 3D:} The evolution of the rovibrational transitions of the $H_2$ molecule with all its degrees of freedoms and considering the magnetic field effect on both nuclei and electrons is shown in Fig.~\ref{fig:all cases both nucleus and electron}d. From Fig.~\ref{fig:all cases both nucleus and electron}d, it is noticeable that in addition to shifts and splits of peaks arising from the electronic effects as we already discussed in Fig~\ref{fig:all cases only electron}d, the rotational and rovibrational bands shows  splitting patterns of the only nuclei case shown in Fig~\ref{fig:all cases only nuclei}d. The resultant spectra at T = 0K and T = 1500K can be seen in Figs.~\ref{fig:3D_Num_NE_T0K} and ~\ref{fig:3D_Num_NE_T1500K} respectively. Fig.~\ref{fig:3D-resolved-te-T=0K} and Fig.~\ref{fig:3D-resolved-os-T=0K} show the evolution of quadrupole allowed transitions and oscillator strengths with magnetic field at T = 0K respectively. We now discuss the evolution of individual peaks. \\

\textit{\textbf{Transitions from ($v=0, \;J=0$), i.e, T = 0K:}} \\
\textbf{Rotational peaks:} The 5 fold degeneracy of the $S_0(0)$ peak at 352, breaks into three groups of peaks as shown in Fig.~\ref{fig:3D-resolved-te-T=0K}a, matching the splitting in Fig.~\ref{fig:all cases only electron}d. Each of the two lower groups split into two sub-branches, which is a signature of the field effect on the nuclei. The oscillator strengths of all the peaks in this family is increased by around 4 orders of magnitude as $B$  changes from $B=0$ to $B=3$ au as shown in Fig.~\ref{fig:3D-resolved-os-T=0K}a.

\textbf{Vibrational peaks:} The $Q_1(2)$ peak at 4140 (at $B=0$) smoothly evolves as $B$ increases as indicated in Fig.~\ref{fig:3D-resolved-te-T=0K}b and Fig.~\ref{fig:3D-resolved-os-T=0K}b commensurate with a decrease in bond length and an increase in bond energy as $B$ increases. The oscillator strength increases by two orders of magnitude.

\textbf{Rovibrational peaks:} The evolution of the $S_1(0)$ peak at 4476 (at $B=0$) reflects that of $S_0(0)$, i.e., it is 5-fold degenerate at $B=0$ and for $B > 0.4$ au, it branches out to three groups of peaks, with the two groups further splitting, obeying the cyclotron coupling and splitting as discussed in the only electron and only nucleus case. However, an interesting feature to note is that the oscillator strength shows oscillatory behavior and a continuously decreasing profile as indicated in Fig.~\ref{fig:3D-resolved-os-T=0K}c. It decreases by 2 to 4 orders, and the transition can be considered allowed only up to $B=1.5$ au above which it becomes disallowed. This trend conveys that as $B$ increases, the $S_1(0)$ and $S_0(0)$ peaks become decoupled. This may be called field-mediated decoupling. 

\textbf{Magnetic field induced rotation/libration peaks:} At B = 0, the peaks at 703 [(0,0) $\rightarrow$ (0,3)], 1168 [(0,0) $\rightarrow$ (0,4)], and 1740 [(0,0) $\rightarrow$ (0,5)] which are quadrupole forbidden get progressively allowed at field strengths of 0.1, 0.2 and 0.3, respectively, and all of them blue shift smoothly up to $B=1.3$ au and then become disallowed as shown in Fig.~\ref{fig:3D-resolved-te-T=0K}d. In the allowed regime majority of them show their corresponding groups splittings through cyclotron coupling, indicating the effect of the field on the nuclei. Their corresponding oscillator strengths increase, as shown in Fig.~\ref{fig:3D-resolved-os-T=0K}d. These may be identified as field induced rotational transitions.

\textbf{Magnetic field-induced rovibrational couplings/peaks:} The $S_0(1)$ at 4250 at $B=0$, gets allowed for $B\ge0.1$ and branches out to two different groups and blue shifts smoothly. The lower group further splits as indicated in Fig.~\ref{fig:3D-resolved-te-T=0K}e. In addition, the peaks at 4798 and 4803 at $B=0$ get allowed for field strengths $0.1\le B \le 0.4$ au. The oscillator strength shows oscillatory and increasing behavior.

\textbf{Magnetic field induced higher vibrational peaks:} The peaks at 5232 and 4942 (most probably corresponding to $S_1(3)$ and $S_1(2)$ at $B=0$), become allowed at higher field strengths $0.8 < B < 2.2$. Their oscillator strengths increase by around 8 orders in this range. An interesting characteristic to note is that, initially the oscillatory strengths increase non-linearly, but for $B > 1.3$ au they almost saturate as can be noticed in Fig.~\ref{fig:3D-resolved-os-T=0K}.

\textit{\textbf{Transitions from ($v=0, \;J \ge 0$), i.e, T=1500 K:}}

\begin{figure}[H]
    \centering
    \centering
    \includegraphics[width=0.92\textwidth]{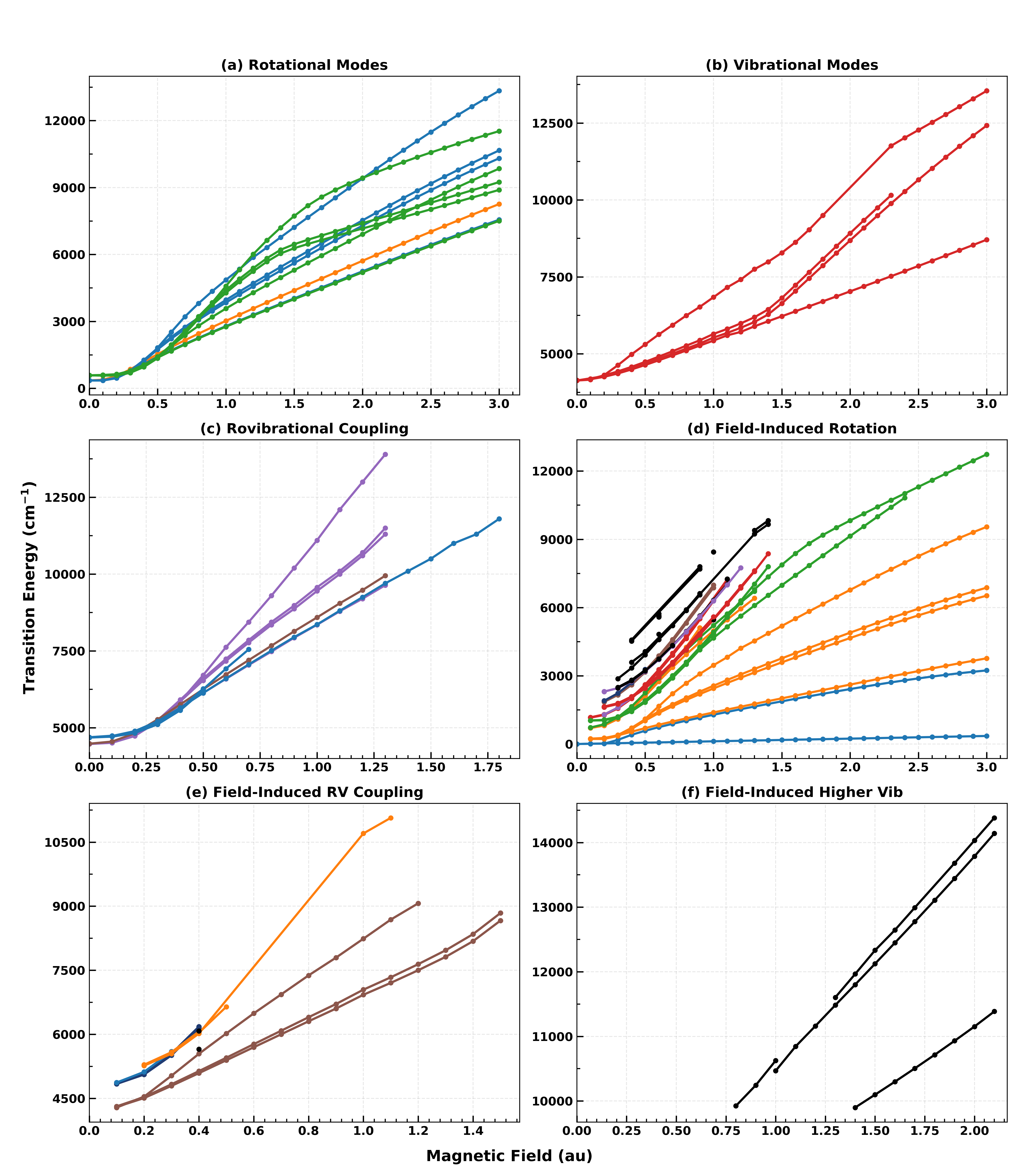}
\caption{Magnetic-field dependence of transition energies at T = 1500K for the 3D electronic PES of H₂ with coupled electron–nuclear motion: (a) rotational, (b) vibrational, (c) rovibrational coupling, (d) field-induced rotational, (e) field-induced rovibrational coupling, and (f) field-induced higher vibrational modes. Transition energies are shown as a function of magnetic field (au).}

    \label{fig:3D-resolved-te-T=1500K}
\end{figure}

\begin{figure}[H]
    \centering
    \centering
    \includegraphics[width=0.92\textwidth]{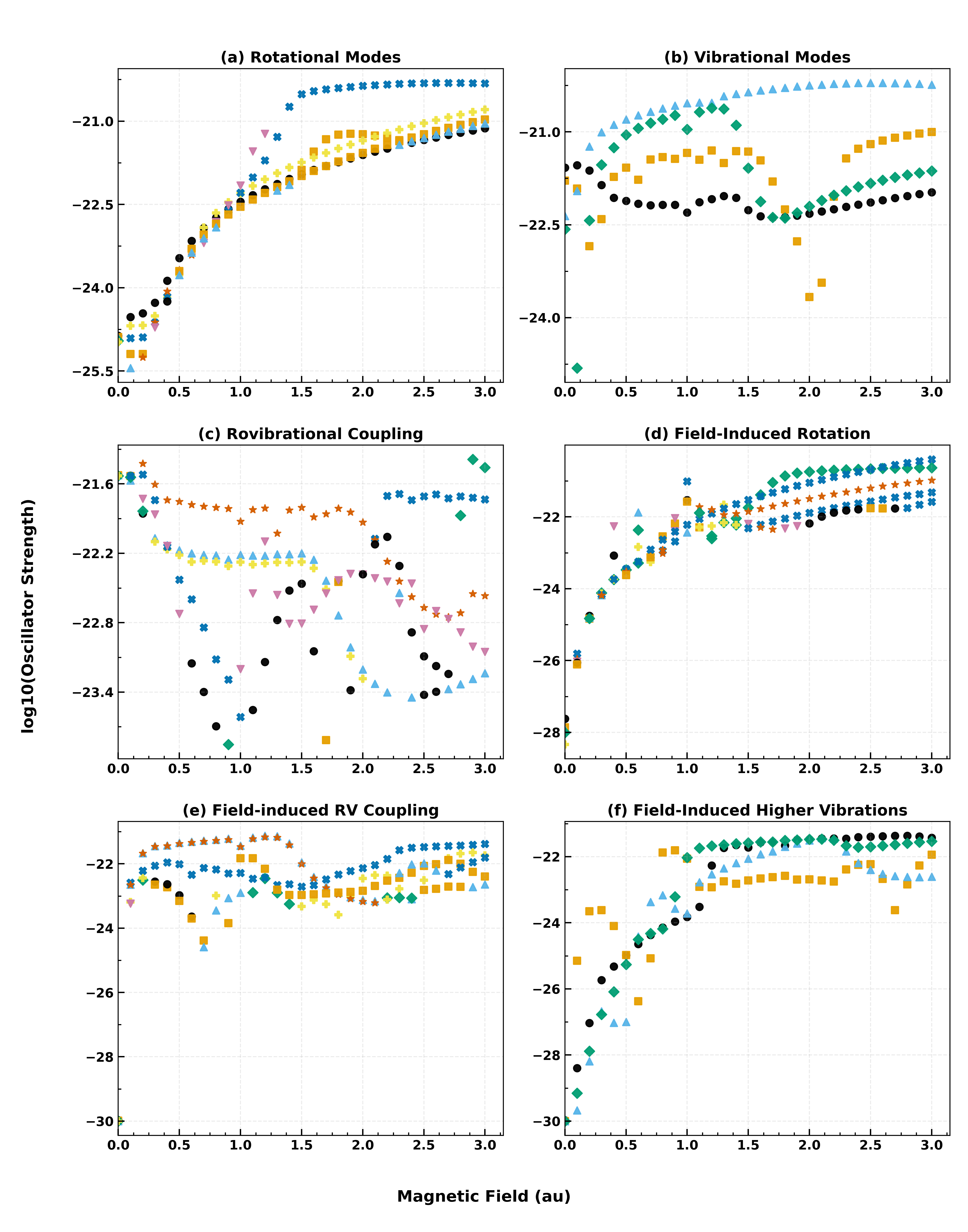}
\caption{Magnetic-field dependence of oscillator strengths at T = 1500K for the 3D electronic PES of H₂ with coupled electron–nuclear motion: (a) rotational, (b) vibrational, (c) rovibrational coupling, (d) field-induced rotational, (e) field-induced rovibrational coupling, and (f) field-induced higher vibrational modes. Transition energies are shown as a function of magnetic field (au), with color grouping indicating low-field spectral proximity.}

    \label{fig:3D-resolved-os-T=1500K}
\end{figure}

On considering $T=1500 K$, along with the transitions from the ground state $(0,0)$, the transitions from higher rotational levels of $v=0$ are also activated. In this subsection we shall focus on the evolution of transitions from the $(0,1)$ level. The evolution of transition energies and oscillator strengths for $T=1500K$, can be seen in Fig.~\ref{fig:3D-resolved-te-T=1500K} and Fig.~\ref{fig:3D-resolved-os-T=1500K} respectively.

\textbf{Rotational peaks:} In addition to the $S_0(0)$ peak at 352, we obtain the $S_0(1)$ at 581 (experimental value is 587). The $S_0(1)$ has seven fold degeneracy. It evolves into four groups among which the three lower groups further split as shown in Fig.~\ref{fig:3D-resolved-te-T=1500K}a. In Fig.~\ref{fig:3D-resolved-os-T=1500K}, the blue and red indicate transitions from two different levels (v,J) = (0,0) and (0,1) respectively. The oscillator strengths of the $S_0(1)$  and $S_0(0)$ peaks evolve in two different groups with the  values increasing roughly by 4 orders of magnitude.

\textbf{Vibrational peaks:} The vibrational peak $Q_1(3)$ at 4134 (experimental value of 4125) is 3-fold degenerate at $B=0$. The degeneracy sustains more or less upto $B=1.3$ au and after that the peak splits into groups of two and one. The upper group further splits (through cyclotron coupling) as shown in Fig.~\ref{fig:3D-resolved-te-T=1500K}b. We can note that, its oscillator strength increases by 2 to 3 orders upto $B=1.3$ au and then has a reverse bump (decrease and then increase) between $1.5\le B \le 2.3$ and saturates thereafter as shown in Fig.~\ref{fig:3D-resolved-os-T=1500K}b.

\textbf{Rovibrational peaks:} $S_1(2)$ becomes accessible at T=1500K and $B=0$. It has 7-fold degeneracy which is a reflection of the coupling of $S_0(2)$ with the vibrational level. It is allowed at low and intermediate fields in the range of $0 < B < 1.8$ as shown in Fig.~\ref{fig:3D-resolved-te-T=1500K}c. Their respective oscillator strengths fluctuate and roughly decrease as the $B$ decreases as shown in Fig.~\ref{fig:3D-resolved-os-T=1500K}c. 

\textbf{Magnetic field induced rotational peaks:} One of the rotations is a pure cyclotron rotation with $12 \;\mathrm{cm^{-1}}$ increment for every $0.1$ au of $B$ increase. Many rotations get allowed at various field strengths - 235, 1037 (5-fold degenerate),  1630, 2306 (7-fold degenerate), and others. The field strengths of $B = 0.6, 0.9, 1.3$ and 1.4 are interesting to note because some rotations are active at these field strengths only.

\textbf{Magnetic field-induced rovibrational peaks:} The $S_1(3)$ line at 5112 gets allowed for $0.1\le B \le 0.4$. It is, in principle, $7-\text{fold}$ degenerate but since we considered only 120 energy levels, we captured only 4 of the levels as shown in Fig.~\ref{fig:3D-resolved-te-T=1500K}e. Their oscillator strengths show an oscillatory pattern in Fig.~\ref{fig:3D-resolved-os-T=1500K}e which eventually increase by up to 3 orders of magnitude. 

\textbf{Magnetic field induced higher vibrational peaks:} At higher fields, $1.3 \le B \le 2.0$, two additional excitations become allowed, the evolution of transition energy and oscillator strengths are shown in Fig.~\ref{fig:3D-resolved-te-T=1500K}f and Fig.~\ref{fig:3D-resolved-os-T=1500K}f, respectively.

\subsection{Comparative Analysis of Classical and Quantum Spectra in External Magnetic Fields}

\begin{table}[htbp]
\centering
\caption{Selected transition energies (cm$^{-1}$) extracted directly from the full dataset without modification, highlighting key spectral features and comparison with Refs.~\citenum{AIMD2,MDLM-1}. HF=Hartree-Fock PES, MP2=Moller-Plesset 2nd order Perturbation Theory PES, CCSD=Coupled Cluster Singles Doubles PES, MD=Molecular Dynamics for Nuclei (classical), QM=Wilson Hamiltonian Diagonalization for Nuclei (quantum mechanical)}
\label{tab:key_transitions_clean}

\small
\setlength{\tabcolsep}{5pt}
\renewcommand{\arraystretch}{1.1}

\begin{tabular}{|c|c|c|c|c|c|}
\hline
\textbf{B (au)} & \textbf{HF + MD} & \textbf{MP2 + MD} & \textbf{HF + QM} & \textbf{MP2 + QM} & \textbf{CCSD + QM} \\
\hline

0 
& 325  & 680   & 361  & 358  & 352 \\
& 4200 & 3760  & 4354 & 4275 & 4140 \\
& 4850 & 5170  & 4698 & 4615 & 4473 \\
\hline

0.1 
& 325  &        & 361  & 358  & 352 \\
&      &        & 360  & 357  & 351 \\
&      &        & 1192 & 1181 & 1161 \\
& 4200 &        & 4393 & 4424 & 4287 \\
& 4850 &        & 5085 & 4999 & 4850 \\
\hline

1 
& 1400 & 1350$\pm$550 & 1372 & 1397 & 1399 \\
& 1500 & 2800,4200    & 1491 & 1516 & 1518 \\
& 2900 &              & 2630 & 2680 & 2685 \\
& 4750 &              & 2733 & 2783 & 2787 \\
& 4850 &              & 2970 & 3020 & 3024 \\
& 6100 &              & 3755 & 3832 & 3842 \\
& 7750 &              & 3873 & 3950 & 3961 \\
& 7850 &              & 4729 & 4840 & 4861 \\
&      & 6000,7500    & 6016 & 6941 & 6927 \\
\hline

\end{tabular}
\end{table}

\begin{figure}
    \centering
    \includegraphics[width=0.98\linewidth]{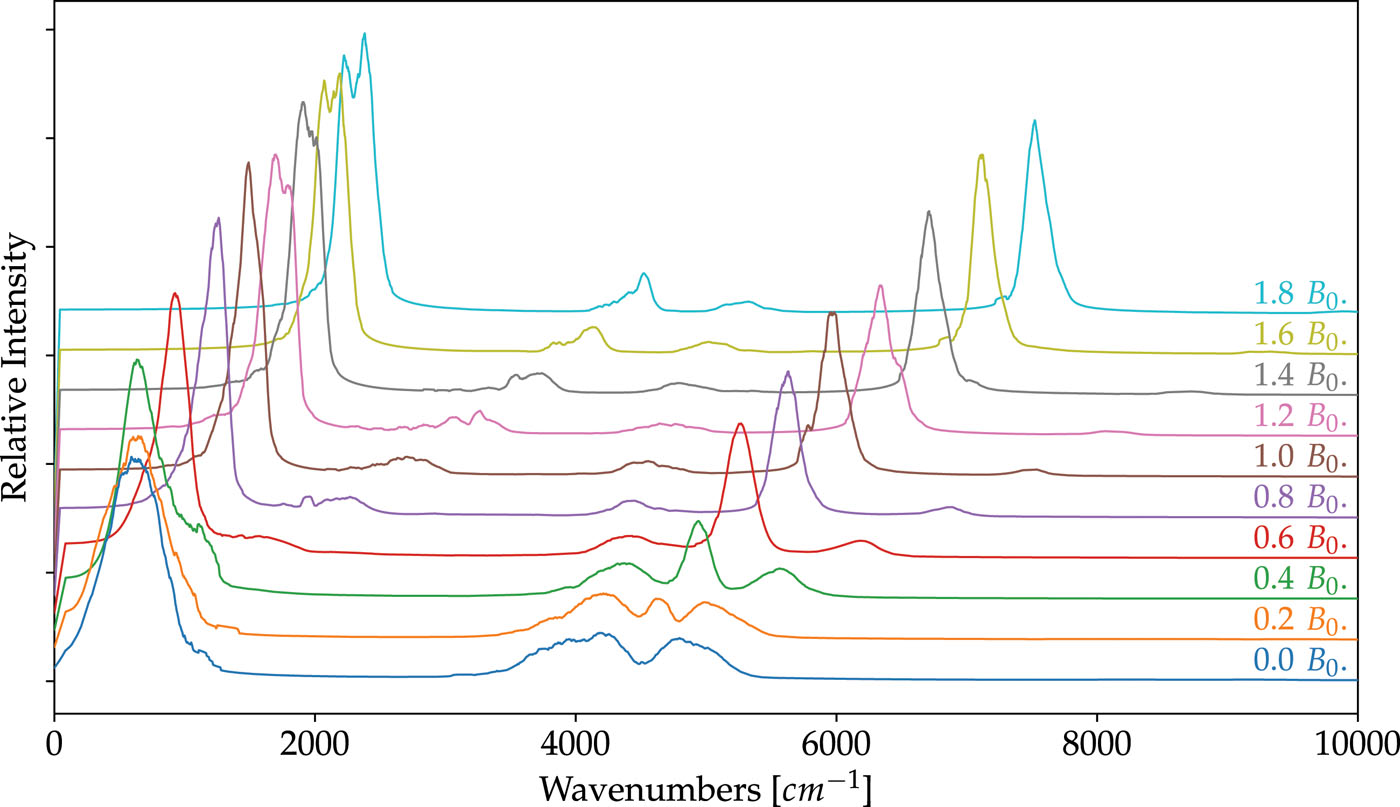}
    \caption{Canonical ensemble spectra from classical MD simulations reproduced from Fig. 6 of (J. Chem. Phys. 7 August 2022; 157 (5): 054106)}
    \label{fig:Erik's spectra}
\end{figure}

\begin{table}[htbp]
    \centering
    \small
    \renewcommand{\arraystretch}{1.2}
    \caption{Classical MD Spectra (J. Chem. Phys. 7 August 2022; 157 (5): 054106): Cluster-based spectral interpretation across magnetic fields $B = 0$--$1.8$ a.u. Note the emergence of discrete states in the Mid-E region as $B$ increases.}
    \label{tab:D1_grid}
    \begin{tabularx}{\textwidth}{@{} c >{\raggedright\arraybackslash}X >{\raggedright\arraybackslash}X >{\raggedright\arraybackslash}X @{}}
        \toprule
        \textbf{B (a.u.)} & \textbf{Low-E Region} & \textbf{Mid-E Region} & \textbf{High-E Region} \\
        \midrule
        0   & $600\pm300$  & ---               & 4200, 4800 \\
        0.2 & $650\pm350$  & weak (3200)       & 4250, 4900 \\
        0.4 & $750\pm400$  & 3500              & 4400, 5100, 5600 \\
        0.6 & $950\pm450$  & 3800              & $5300\pm300$, 6200 \\
        0.8 & $1150\pm500$ & 4000              & 5650, 6800 \\
        1.0 & $1350\pm550$ & 2800, 4200        & 6000, 7500 \\
        1.2 & $1600\pm600$ & 3300, 4300        & 6350, 7800 \\
        1.4 & $1950\pm500$ & 3500, 3800, 4400  & 6700, 7200 \\
        1.6 & $2250\pm450$ & 4100, 4600, 5300  & 7150, 7500 \\
        1.8 & $2400\pm500$ & 4500, 5400, 5700  & 7550, 8100 \\
        \bottomrule
    \end{tabularx}
\end{table}

\vspace{1em} 

\begin{table}[htbp]
    \centering
    \small
    \renewcommand{\arraystretch}{1.2}
    \caption{This work: Raw spectral distribution. Data represented as mean energies ($\mu$), spreads ($\sigma$), and observed spectral ranges in parentheses.}
    \label{tab:D2_grid}
    \begin{tabularx}{\textwidth}{@{} c >{\raggedright\arraybackslash}X >{\raggedright\arraybackslash}X >{\raggedright\arraybackslash}X @{}}
        \toprule
        \textbf{B (a.u.)} & \textbf{Low-E Region ($\mu \pm \sigma$)} & \textbf{Mid-E Region (Range)} & \textbf{High-E Region} \\
        \midrule
        0   & $\sim470\pm120$ (350-600)  & gap (2000-3500)         & $\sim4400\pm250$ \\
        0.2 & $\sim400\pm200$             & onset (2000-3000)       & emerging tail \\
        0.4 & $\sim900\pm400$             & continuous (2000-3500)  & tail up to $\sim4500$ \\
        0.6 & $\sim1200\pm600$            & dense (3000-4500)       & 4500-5500 \\
        0.8 & $\sim1500\pm700$            & dominant band (3000-5000) & 5000-6000 \\
        1.0 & $\sim1800\pm800$            & continuous band          & 5000-6500 \\
        1.2 & $\sim1850\pm450$            & dense (3000-5000)       & $\sim6100\pm500$ \\
        1.4 & $\sim2100\pm500$            & dense (3200-5000)       & $\sim6500\pm600$ \\
        1.6 & $\sim2350\pm550$            & broad (3800-6000)       & $\sim7100\pm700$ \\
        1.8 & $\sim2600\pm600$            & very broad (4200-6500)  & $\sim8500\pm1200$ (tail $>11000$) \\
        \bottomrule
    \end{tabularx}
\end{table}

Classical MD simulation have been carried out by other groups to compute the molecular rovibrational spectra of $^1\mathrm{H}_2$ in strong magnetic fields~\cite{AIMD2,MDLM-1,VDF1}. As discussed before, we have carried out fully quantum mechanical computations. The electronic PES in both cases is computed quantum mechanically. In Refs.~\citenum{AIMD2}, \citenum{AIMD1} and \citenum{VDF1} through the same program package LONDON while Ref.~\citenum{MDLM-1} uses TURBOMOLE. Ref. ~\citenum{AIMD2} and ~\citenum{VDF1} use a HF PES, Ref.~\citenum{MDLM-1} uses MP2 and we use CCSD. We have also recomputed the transition energies using HF and MP2 electronic PESs with the LONDON program package and the comparison of the transition energies is given in Table ~\ref{tab:key_transitions_clean} and the main features of the classical spectra~\cite{MDLM-1} are summarized in Table~\ref{tab:D1_grid}. Note that since the number of transitions for T=0K and T=1500K differ, the average values of the corresponding common transitions differ between Tables. Here we shall compare between the spectra generated through classical MD simulation in a magnetic field~\cite{AIMD2,MDLM-1}, Fig.~\ref{fig:Erik's spectra} and our spectra at T=1500 K, Fig.~\ref{fig:2D_Num_NE_T1500K}. Importantly, neither approach is universally superior; rather, they provide distinct but overlapping perspectives on the underlying physics.

In classical dynamics, the Lorentz force acting on charged particles was incorporated into the equations of motion, leading to field-driven trajectories. The nuclei were propagated using the Tajima propagator, which was found to perform best for magnetic-field-driven systems. The Berry curvature is computed from the electronic wave function and used to screen the nuclear charge from the field. The absence of screening leads to the appearance of a low-frequency feature associated with cyclotron motion ($\omega = 120 \;\mathrm{cm}^{-1}$ at $\mathrm{B}=1$ au). Other than this, the effect of Berry screening on the rovibrational spectra has been shown to be small~\cite{AIMD2, Non-BO-3} for $\mathrm{H}_2$ and is not included in our computations and does not significantly alter the dominant spectral features in the field range considered. 

Another distinction arises in the treatment of the center-of-mass (COM) motion. 
The quantum calculations are performed in the center-of-mass frame, where the internal and external degrees of freedom are cleanly separated for $\mathrm{H}_2$ as discussed in Sec.~\ref{sec:theory}. In contrast, classical MD includes all degrees of freedom in the trajectory evolution, without explicit separation, which allows a mixing of internal and external motions. The classical MD is thus general enough to treat systems where field-mediated coupling of COM and internal motion occurs. Thus, each approach treats COM dynamics differently, with corresponding implications for spectral interpretation.

A key insight provided by these studies is that the splitting of rotational modes cannot be attributed solely to the anisotropy of the PES. Instead, it arises from velocity-dependent forces associated with Berry curvature, 
which acts as an effective gauge field in nuclear configuration space. The Berry curvature introduces gyroscopic couplings that break time-reversal symmetry and lift the degeneracy between clockwise and counterclockwise rotational motion. As a result, rotational eigenmodes acquire complex character, corresponding physically to precessional motion rather than simple oscillations. 

The MD spectra employs a dipolar (P-, Q-, R-branch) classification, which, while useful for organizing spectral features, is not applicable to the homonuclear $\mathrm{H_2}$ molecule due to the absence of a permanent dipole moment. Peaks appearing in the experimental IR spectra of $\mathrm{H}_2$ at $\mathrm{B}=0$ are quadrupolar. Even in the presence of the uniform magnetic field, inversion symmetry is retained with $\mathrm{C}_i$ being the lowest symmetry point group for the $\mathrm{H}_2$+field system~\cite{Pausch2021MolecularFields}. In the quantum-dynamical spectra we have computed the quadrupolar transition moment integral, providing a formally consistent description of allowed transitions. It is important to remember that the intensities in classical MD spectra represent a probability distribution of sampled trajectories which is critically dependent on initial conditions, propagation times, and other user-defined parameters. The quantum approach is a true reflection of relative peak intensities. 

While the quantum spectra provide a more detailed representation of the density of states and transition pathways, this increased resolution can also make interpretation more complex. The classical spectra, by contrast, effectively filters the spectral response to its most prominent components, which can be advantageous for identifying the dominant physics. However for truly computing spectra for astrochemical interpretation, the quantum framework is essential.

In what follows, the comparison between the classical MD and quantum spectra is organized in terms of similarities and differences across low-, intermediate-, and strong-field regimes. The key features are summarized in Tables ~\ref{tab:D1_grid} and ~\ref{tab:D2_grid}.

\textbf{Low-field regime ($B \lesssim 0.2$ a.u.):} In the low-field regime, both approaches converge to a near field-free molecular description, characterized by well-resolved, discrete (singlet-like) transitions. The low-energy region lies in the range $\sim400$-$650$ (this work: $\sim470\pm120$ cm$^{-1}$; Ref.~\citenum{MDLM-1}: $600\pm300$), while high-energy vibrational features appear around $4200$-$4800$. A pronounced spectral gap of $\sim2000$-$3500$ separates rotational and vibrational excitations, indicating weak coupling between these degrees of freedom. Both approaches exhibit consistent blue shifting with increasing magnetic field, reflecting the onset of magnetic confinement and mild stiffening of the vibrational mode. In this regime, classical trajectories and quantum eigenstates remain closely aligned, and the spectral structure is dominated by isolated transitions with minimal mixing.

\textbf{Intermediate-field regime ($0.4 \lesssim B \lesssim 1.3$ a.u.):} In the intermediate-field regime, both datasets capture the onset of magnetic-field-–induced coupling between rotational, vibrational, and transverse (Landau-like) degrees of freedom. The intermediate-field regime reveals the most significant differences, reflecting the increasing complexity of the underlying dynamics. The energy scales shift consistently: the low-energy region evolves from $\sim700$ to $\sim1800$, while high-energy features shift from $\sim4400$-$5600$ to $\sim6000$-$7500$. Both spectra exhibit the emergence of mid-energy features ($\sim2000$-$5000$), associated with magnetic-field-induced rotational motion and rovibrational coupling. Spectroscopically, isolated singlets evolve into multiplet structures (doublets and triplets), reflecting degeneracy lifting and level repulsion. Thus, both approaches consistently capture the transition toward increased state mixing and spectral complexity.

Spectroscopically, the classical spectra retain discrete peak structures with identifiable branches, whereas the quantum spectra reveal a dense manifold of transitions spanning $\sim2000$-$5000$. Importantly, this difference should not be viewed solely as a limitation of the classical approach: rather, the classical spectra provides a clearer identification of dominant frequency components, while the quantum spectra resolves a larger number of weaker transitions arising from state mixing.

Additionally, while both approaches exhibit overall blue shifting with increasing field strength, the quantum spectra display more intricate evolution, including intensity modulations associated with wavefunction changes, appearance, disappearance and merging of peaks as well as clear signatures of degeneracy lifting, avoided crossings and state mixing, whereas the classical spectra tend to exhibit smoother, monotonic shifts and splittings.

\textbf{Strong-field regime ($B \gtrsim 1.3$ a.u.):} In the strong-field regime, both datasets reflect a magnetically dominated spectral structure. The low-energy region lies in the range $\sim1600$-$2600$, while the mid-energy region spans $\sim3000$-$6000$. Both approaches show continued blue shifting and an increase in spectral density, consistent with strong magnetic confinement and enhanced coupling between degrees of freedom. The clustering of spectral features indicates the breakdown of simple quantum number assignments and the emergence of a dense manifold of states.

In the strong-field regime, the divergence between the two approaches becomes more pronounced. The classical spectra continue to display discrete peaks with relatively limited bandwidth ($\sim1000$-1500), providing a compact representation of the dominant dynamical frequencies. In contrast, the quantum spectra exhibit a much denser set of transitions extending to $\sim8500\pm1200$ with a high-energy tail beyond $11000$, corresponding to an effective bandwidth of $\sim3000$-6000. This difference likely due to the lack of sufficiently high initial velocities in the classical MD simulations.

\section{Summary and Future Outlook}

\begin{figure}[H]
    \centering
    \includegraphics[width=0.7\linewidth]{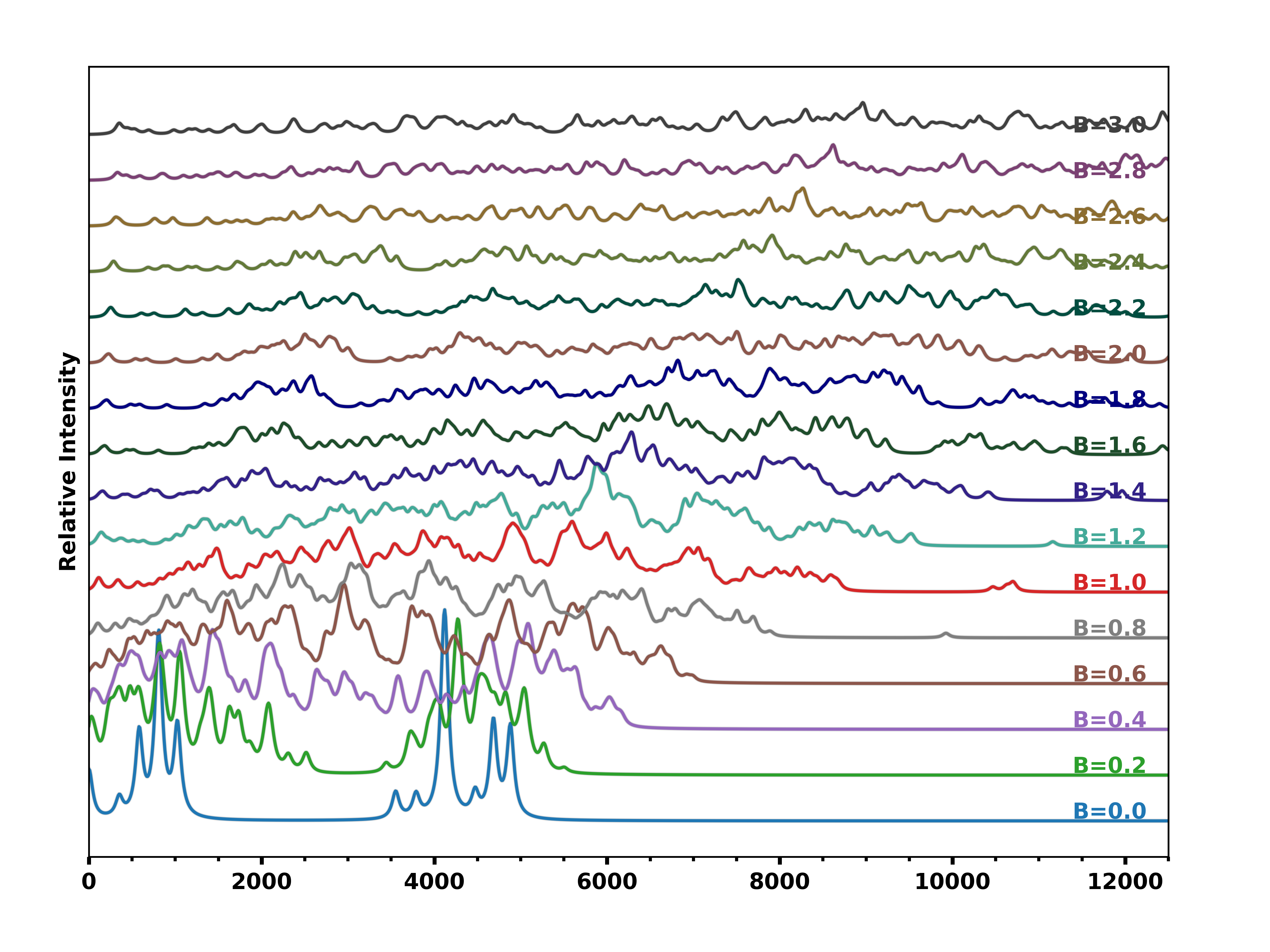}
    \caption{Electronic Potential (Magnetic field acting on both nuclei and electrons) : Rovibrational Line Spectra of $H_2$ subjected to  magnetic field at $T = 12000$ K}
    \label{fig:2D_Num_NE_T12000K}
\end{figure}

This work presents the first quantum-mechanical investigation of the electric quadrupolar rovibrational spectra of molecular hydrogen in external magnetic fields. Within a grid-based Hamiltonian framework incorporating magnetic effects via Peierls substitution, the evolution of energy levels, transition energies, and oscillator strengths has been explored across different field regimes. The accuracy of the method has been validated against high-precision experimental spectra at $B = 0$ and against analytical values for a 2D harmonic oscillator potential with a perpendicular magnetic field~\cite{jctc1}. Comparison against earlier classical MD simulations have also revealed similar dominant physics with our results being more suitable for comparison with experimental spectra on account of a more accurate PES, full quantum mechanical treatment, better resolution and respecting selection rules. The non-perturbative quantum mechanical treatment ensures that all coupling effects are considered across a wide range of field strengths. A sample of the computed spectra for the surface of a magnetic white dwarf corresponding to a representative temperature of T=12000K is presented in Fig.~\ref{fig:2D_Num_NE_T12000K}. As is obvious, the spectra is dense and convoluted and very little peak characteristics or indication of underlying physics can be deduced from it. Thus, this work has explored various simplified cases, reduced dimensionality and lower temperatures, with particular emphasis on isolating and understanding the distinct roles of nuclear and electronic degrees of freedom. 

A central outcome is the clear differentiation of magnetic-field responses arising from nuclei and electrons. The nuclear contribution follows the expected linear Zeeman behavior for the greater part of the magnetic field range considered (B=0-3 au), with slight quadratic diamagnetic effects showing up for a few rovibrational states beyond 2 au. The electronic contribution, on the other hand, reflects a more intricate competition between paramagnetic and diamagnetic effects, with the PES shifting non-linearly with field from the very beginning, leading to non-linear shifts in transition energies, new couplings and PES-induced splittings of rovibrational levels. When both subsystems are treated simultaneously, the spectra exhibit a richer reorganization, where shifts, splittings, and the redistribution of spectral intensity indicate the onset of coupled dynamics beyond simple additive behavior. The comparative study of the rotations and vibrations in 2D and 3D helps in identifying peaks and isolating field-driven and coupling driven peak evolutions. 

The present framework relies on a deliberate set of approximations. The use of the reduced mass-reduced charge framework enables a tractable quantum treatment for diatomics, but the path to larger molecules is not yet clear. At present, geometric and Berry curvature effects are absent but can be included in the future in a manner similar to that in the classical MD approach. Although such contributions are expected to be small for H$_2$, particularly in the parameter range considered, their inclusion would be necessary for a more complete and systematically improvable description, especially at stronger fields or in systems with enhanced electronic complexity.

Overall, this work provides a quantum mechanical but physically transparent and computationally robust framework for analyzing magnetic-field effects in molecular spectra. By linking detailed numerical results with underlying physical mechanisms, it could provide a foundation for future studies aimed at interpreting astrochemical spectra from magnetic stellar objects.

\section*{Supplementary Information}
Supplementary Information (SI) contains the data corresponding to the figures in the manuscript. Additional information comprising of PES visualizations, energy level plots, and more detailed spectral evolution is also given in the SI.

\section*{Conflict of Interest}
The authors declare no competing financial interest.

\section*{Acknowledgements}
NY would like to acknowledge INSPIRE and the Department of Science and Technology, Govt. of India, for funding this work with an INSPIRE fellowship [Grant No: IF200433] and IISER Kolkata for computation facilities. We acknowledge National Supercomputing Mission (NSM) for providing computing resources of ‘PARAM RUDRA’ at S.N. Bose National Centre for Basic Sciences, which is implemented by C-DAC and supported by the Ministry of Electronics and Information Technology (MeitY) and Department of Science and Technology (DST), Government of India. NY would like to thank his colleagues -  Pratyush Bhattacharjya and Urvi Mukherjee, for motivating him to shift his code from FORTRAN-based to Python-based, which greatly enhanced computational efficiency. 



\newpage

\bibliography{source.bib}
\end{document}